\documentstyle[12pt]{article}
\input{epsf}

\def\hybrid{\topmargin -20pt	\oddsidemargin 0pt
	\headheight 0pt	\headsep 0pt
        \textwidth 6.35in
        \textheight 9.65in
	\marginparwidth .875in
	\parskip 5pt plus 1pt	\jot = 1.5ex}

\catcode`\@=11
\newcount\hour
\newcount\minute
\newtoks\amorpm
\hour=\time\divide\hour by60
\minute=\time{\multiply\hour by60 \global\advance\minute by-\hour}
\edef\standardtime{{\ifnum\hour<12 \global\amorpm={am}%
	\else\global\amorpm={pm}\advance\hour by-12 \fi
	\ifnum\hour=0 \hour=12 \fi
	\number\hour:\ifnum\minute<10 0\fi\number\minute\the\amorpm}}
\edef\militarytime{\number\hour:\ifnum\minute<10 0\fi\number\minute}

\def\marginnote#1{}
\def\draftlabel#1{{\@bsphack\if@filesw {\let\thepage\relax
   \xdef\@gtempa{\write\@auxout{\string
      \newlabel{#1}{{\@currentlabel}{\thepage}}}}}\@gtempa
   \if@nobreak \ifvmode\nobreak\fi\fi\fi\@esphack}
	\gdef\@eqnlabel{#1}}
\def\@eqnlabel{}
\def\@vacuum{}
\def\draftmarginnote#1{\marginpar{\raggedright\scriptsize\tt#1}}

\def\draft{\oddsidemargin -.2truein
	\def\@oddfoot{\sl preliminary draft \hfil
	\rm\thepage\hfil\sl\today\quad\militarytime}
	\let\@evenfoot\@oddfoot	\overfullrule 3pt
	\let\label=\draftlabel
	\let\marginnote=\draftmarginnote
   \def\@eqnnum{(\theequation)\rlap{\kern\marginparsep\tt\@eqnlabel}%
\global\let\@eqnlabel\@vacuum}  }

\def\preprint{\twocolumn\sloppy\flushbottom\parindent 2em
	\leftmargini 2em\leftmarginv .5em\leftmarginvi .5em
	\oddsidemargin -.5in	\evensidemargin -.5in
	\columnsep .4in	\footheight 0pt
	\textwidth 10.in	\topmargin  -.4in
	\headheight 12pt \topskip .4in
	\textheight 6.9in \footskip 0pt
	\def\@oddhead{\thepage\hfil\addtocounter{page}{1}\thepage}
	\let\@evenhead\@oddhead	\def\@oddfoot{}	\def\@evenfoot{} }
\def\numberbysection{\@addtoreset{equation}{section}
	\def\theequation{\thesection.\arabic{equation}}}

\def\underline#1{\relax\ifmmode\@@underline#1\else
	$\@@underline{\hbox{#1}}$\relax\fi}

\def\titlepage{\@restonecolfalse\if@twocolumn\@restonecoltrue
\onecolumn
     \else \newpage \fi \thispagestyle{empty}\c@page\z@
	\def\thefootnote{\fnsymbol{footnote}} }

\def\endtitlepage{\if@restonecol\twocolumn \else \newpage \fi
	\def\thefootnote{\arabic{footnote}}
	\setcounter{footnote}{0}}  

\catcode`@=12
\relax
\hybrid
\begin{document}

\def\be{\begin{equation}}
\def\ee{\end{equation}}
\def\bea{\begin{eqnarray}}
\def\eea{\end{eqnarray}}
\def\tfrac#1#2{{\textstyle{#1\over #2}}}
\def\r#1{{(\ref{#1})}}
\def\half{\tfrac{1}{2}}
\def\quart{\tfrac{1}{4}}
\def\nn{\nonumber}
\def\na{\nabla}
\def\ms{M_{\rm string}}
\def\gs{g_{\rm string}}
\def\pd{\partial}
\def\a{\alpha}
\def\b{\beta}
\def\g{\gamma}
\def\d{\delta}
\def\m{\mu}
\def\n{\nu}
\def\t{\tau}
\def\l{\lambda}
\def\mh{\hat\m}
\def\nh{\hat\n}
\def\rh{\hat\rho}
\def\th{\vartheta}
\def\s{\sigma}
\def\e{\epsilon}
\def\ap{\alpha'}
\def\al{\alpha}
\font\mybb=msbm10 at 12pt
\def\bb#1{\hbox{\mybb#1}}
\def\Z{\bb{Z}}
\def\R{\bb{R}}
\def\Z{\bb{Z}}
\def\R{\bb{R}}
\def\C{\bb{C}}
\def\square{\hbox{{$\sqcup$}\llap{$\sqcap$}}}
\def\dslash{{\partial\hspace{-7pt}/}}
\hyphenation{re-pa-ra-me-tri-za-tion}
\hyphenation{trans-for-ma-tions}
\begin{titlepage}
\begin{center}
\hfill CERN-TH/97-217\\
\hfill hep-th/9708130\\

\vskip 2cm
{\Large \bf INTRODUCTION  TO  NON-PERTURBATIVE STRING THEORY}
\vskip .8in

{\bf Elias Kiritsis}\footnote{e-mail: KIRITSIS@NXTH04.CERN.CH}
\vskip .1in
{\em  Theory Division,  CERN, CH-1211 \\
      Geneva 23, SWITZERLAND}
\end{center}
\vskip 1in
\begin{abstract}
A brief introduction to the non-perturbative structure of string theory
is presented. Various non-perturbative dualities in ten and six dimensions as well as D-branes are discussed.

\end{abstract}
\vskip 3cm

{\tt Lectures presented at the CERN-La Plata-Santiago de Compostella
School
of Physics, La Plata, May 1997.}
\vskip 2cm
\begin{flushleft}
CERN-TH/97-217\\
May 1997\\
\end{flushleft}

\end{titlepage}

\renewcommand{\thepage}{\arabic{page}}
\setcounter{page}{1}
\setcounter{footnote}{1}
\renewcommand{\theequation}{\thesection.\arabic{equation}}
\tableofcontents
\newpage

\renewcommand{\theequation}{\thesection.\arabic{equation}}
\section{Non-perturbative string dualities: a foreword\label{nonpert}}
\setcounter{equation}{0}

In these lectures I will give a brief guide to some recent developments
towards understanding the non-perturbative aspects of string theories.
There  was a parallel developement in the context
of supersymmetric field theories, \cite{s,sw}.
We will not discuss here the field theory case. The interested reader
may consult several comprehensive review articles \cite{n1,nn2}.
We would point out however that the field theory non-perturbative
dynamics is naturally understood in the context of string theory
and there was important cross-fertilization between the two
disciplines.

In ten dimensions there are five distinct consistent
supersymmetric string theories, type-IIA,B, heterotic
(O(32),E$_8\times$E$_8$)
and the unoriented O(32) type I theory that contains also open strings.
The two type-II theories have N=2 supersymmetry while the others only
N=1.
An important question we would like to address is: Are these strings
theories
different or just different aspects of the same theory?

\begin{figure}
\begin{center}
\leavevmode
\epsfbox{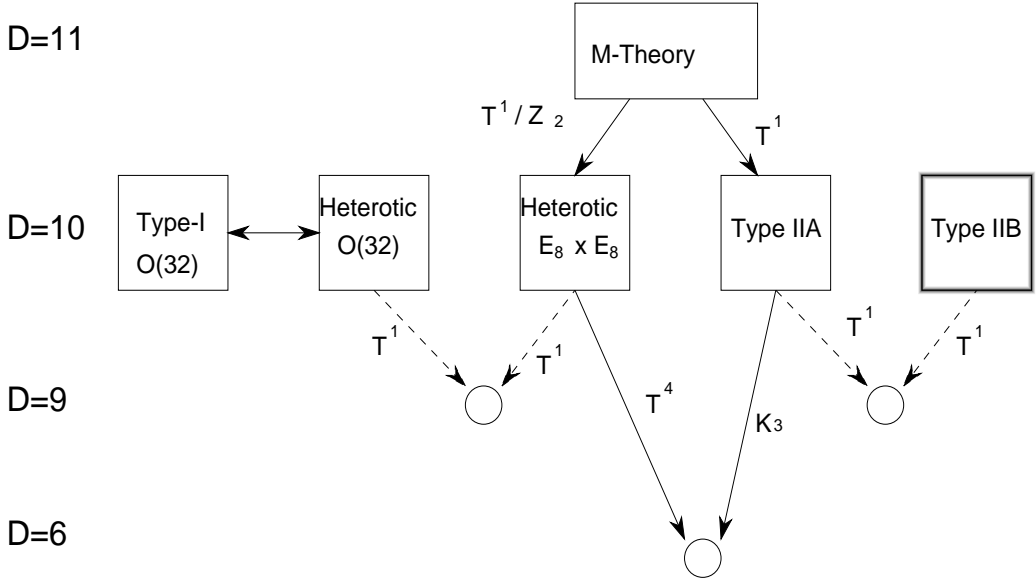}
\caption[]{The Web of duality symmetries between string theories.
Broken lines
correspond to perturbative duality connections. Type-IIB in ten
dimensions  is supposed
to be self-dual under SL(2,$\Z$).}
\label{f19}
\end{center}\end{figure}

In fact, by compactifying one dimension on a circle we can show that we
can connect the two heterotic theories as well as the two type-II
theories.
This is schematically represented with the broken arrows in Fig.
\ref{f19}.

We will first show how the heterotic O(32) and E$_8\times$E$_8$
theories are connected in $D=9$.
Upon compactification on a circle of radius $R$ we can also turn on 16
Wilson lines.
The partition function of the O(32) heterotic theory then can be
written as
\be
Z^{\rm O(32)}_{D=9}={1\over (\sqrt{\tau_2}\eta\bar\eta)^7}
{\Gamma_{1,17}(R,Y^I)\over
\eta\bar\eta^{17}}~{1\over
2}\sum_{a,b=0}^1~(-1)^{a+b+ab}~{\th^4[^a_b]\over \eta^4}
\,,\label{557}\ee
where the lattice sum $\Gamma_{p,p+16}$ is
\be
Z_{p,p+16}(G,B,Y)={\sqrt{{\rm det~G}}\over
\t_2^{p/2}\eta^p\bar\eta^{p+16}}
\sum_{m^{\a},n^{\a}\in Z}\exp\left[-{\pi\over \t_2}(G+B)_{\a\b}
(m^{\a}+\t n^{\a})(m^{\b}+\bar\t n^{\b})\right]\times
\label{B2}\ee
$$\times {1\over
2}\sum_{a,b=0}^1~\prod_{I=1}^{16}~e^{i\pi(m^{\a}Y^I_{\a}
Y^I_{\b}n^{\b}-b~n^{\a}Y^I_{\a})}~\bar\th\left[^{a-2n^{\a}Y^I_{\a}}_
{b-2m^{\b}Y^I_{\b}}\right]
$$

We will focus on some special values for the Wilson lines $Y^I$, namely
we will take eight among them to be zero and the other eight to be 1/2.
Then, the lattice sum (in Lagrangian representation) can be rewritten
as
$$
\Gamma_{1,17}(R)=R\sum_{m,n\in Z}\exp\left[-{\pi R^2\over \t_2}|m+\t
n|^2\right]{1\over 2}\sum_{a,b}~\bar\th^8[^a_b]~\bar\th^8[^{a+n}_{b+m}]
$$
\be={1\over 2}\sum_{h,g=0}^1 \Gamma_{1,1}(2R)[^h_g]~{1\over
2}\sum_{a,b}~\bar\th^8[^a_b]~\bar\th^8[^{a+h}_{b+g}]
\,,\label{558}\ee
where $\Gamma_{1,1}[^h_g]$ are the $Z_2$ translation blocks of the
circle partition function
\be
\Gamma_{1,1}(R)[^h_g]=R\sum_{m,n\in Z}\exp\left[-{\pi R^2\over \t_2}
\left|\left(m+{g\over 2}\right)+\tau\left(n+{h\over
2}\right)\right|^2\right]
\label{559}\ee
\be
={1\over R}\sum_{m,n\in Z}~(-1)^{mh+ng}~\exp\left[-{\pi\over \t_2
R^2}|m+\t n|^2\right]
\,.\label{560}\ee
In the $R\to \infty$ limit (\ref{559}) implies that only $(h,g)=(0,0)$
contributes in the sum in (\ref{558}) and we end up with the O(32)
heterotic string in ten dimensions.
In the $R\to 0$ limit the theory decompactifies again, but from
(\ref{560})
we deduce that all $(h,g)$ sectors contribute equally in the limit.
The sum on $(a,b)$ and $(h,g)$ factorizes and we end up with the
E$_8\times$ E$_8$ theory in ten dimensions.
Both theories are different limiting points (boundaries) in the moduli
space of toroidally compactified heterotic strings.

In the type-II case the situation is similar.
We compactify on a circle. Under an $R\to 1/R$ duality
\be
\pd X^9\to \pd X^9\;\;\;,\;\;\;\psi^9\to\psi^9\;\;\;,\;\;\;\bar\pd X^9
\to -\bar\pd X^9 \;\;\;,\;\;\; \bar\psi^9\to -\bar\psi^9
\,.\label{561}\ee
Due to the change of sign of $\bar\psi^9$ the projection in the $\bar
R$ sector is reversed.
Consequently the duality maps type-IIA to type-IIB and vice versa.
We can also phrase this in the following manner: The $R\to \infty$
limit of the toroidally compactified type-IIA string gives the type-IIA
theory in ten dimensions. The $R\to 0$ limit gives the type-IIB theory
in ten dimensions.

Apart from these perturbative connections, today we have evidence that
all supersymmetric string theories are connected.
Since they look very different in perturbation theory, the connections
must necessarily involve strong coupling.

First, there is evidence that the type-IIB theory has an SL(2,$\Z$)
symmetry which, among other things, inverts the coupling constant
\cite{sch,ht}.
Consequently, the strong coupling limit of type-IIB is given also by
the weakly-coupled type-IIB theory.
Upon  compactification, this symmetry combines with the
perturbative $T$-duality symmetries to produce a large discrete duality
group known as the $U$-duality group, which is the discretization of
the non-compact continuous symmetries of the maximal effective
supergravity theory.
In table 3 below, the $U$-duality groups are given for various
dimensions.
They were conjecture to be exact symmetries in \cite{ht}.
A similar remark applies to non-trivial compactifications.

\vskip 1cm
\centerline{
\begin{tabular}{|c|c|c|c|}\hline
Dimension&SUGRA symmetry& T-duality&U-duality\\ \hline\hline
10A&SO(1,1,\R)/Z$_2$&{\bf 1}&{\bf 1}\\\hline
10B&SL(2,\R)&{\bf 1}&SL(2,\Z)\\\hline
9&SL(2,\R)$\times$O(1,1,\R)&Z$_2$&SL(2,\Z)$\times$Z$_2$\\\hline
8&SL(3,\R)$\times$SL(2,\R)&O(2,2,\Z)&SL(3,\Z)$\times$SL(2,\Z)\\\hline
7&SL(5,\R)&O(3,3,\Z)&SL(5,\Z)\\\hline
6&O(5,5,\R)&O(4,4,\Z)&O(5,5,\Z)\\\hline
5&E$_{6(6)}$&O(5,5,\Z)&E$_{6(6)}$(\Z)\\\hline
4&E$_{7(7)}$&O(6,6,\Z)&E$_{7(7)}$(\Z)\\\hline
3&E$_{8(8)}$&O(7,7,\Z)&E$_{8(8)}$(\Z)\\\hline
\end{tabular}}
\bigskip

\centerline{{ Table 3}: Duality symmetries for the compactified type-II
string.}
\vskip 1cm

Also, it can be argued that the strong coupling limit of type-IIA
theory
is described by an eleven-dimensional theory named ``M-theory"
\cite{va}.
Its low-energy limit is eleven-dimensional supergravity.
Compactification of M-theory on circle with very small radius gives the
perturbative type-IIA theory.

If instead we compactify M-theory on the $Z_2$ orbifold of the circle
$T^1/Z_2$
then we obtain the heterotic E$_8\times$E$_8$ theory, \cite{ee}.
When the circle is large the heterotic theory is strongly coupled while
for small radius it is weakly coupled.

Finally, the strong coupling limit of the O(32) heterotic string theory
is the type I O(32) theory and vice versa, \cite{h-I}.

There is another non-trivial non-perturbative connection in six
dimensions:
The strong coupling limit of the six-dimensional toroidally
compactified heterotic string is given by the type-IIA theory
compactified on K3 and vice versa \cite{ht}.

\begin{figure}
\begin{center}
\leavevmode
\epsfbox{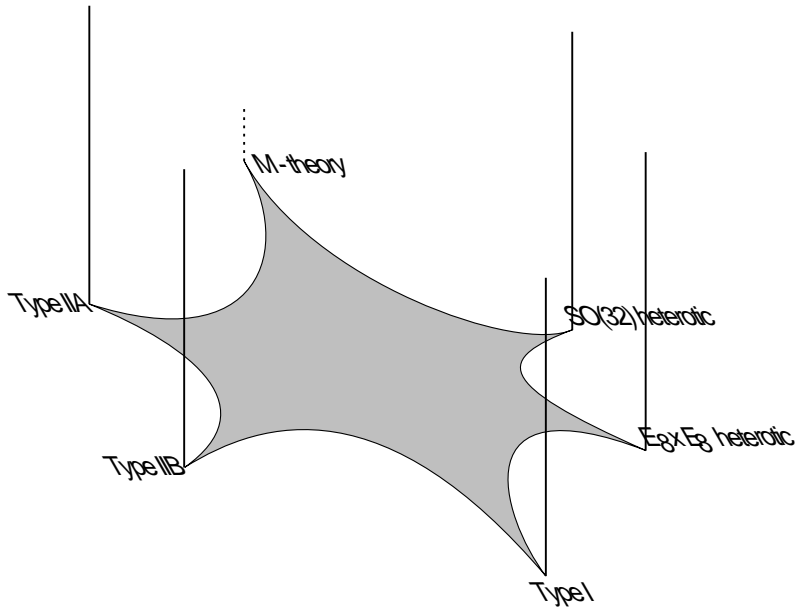}
\caption[]{A unique theory and its various limits.}
\label{f20}
\end{center}\end{figure}

Thus, we are led to suspect that there is an underlying ``universal"
theory
whose various limits in its ``moduli" space produce the weakly coupled
ten-dimensional supersymmetric string theories as depicted in Fig.
\ref{f20}
 (borrowed from \cite{P2}).
The correct description of this theory is unknown although there is a
proposal
that it might have a matrix description \cite{matrix}, inspired from
D-branes
\cite{Dp}, which reproduces
the perturbative IIA string in ten dimensions \cite{m10}.

We will provide with a few more explanations and arguments
supporting the non-perturbative connections mentioned above.
But before we get there, we will need some ``non-perturbative tools",
namely the notion of BPS
states and p-branes, which I will briefly describe.

\renewcommand{\theequation}{\thesection.\arabic{equation}}
\section{Antisymmetric tensors and p-branes\,.\label{antisym}}
\setcounter{equation}{0}

The various string theories have massless
antisymmetric tensors in their spectrum.
We will use the language of forms and we will represent a rank-$p$
antisymmetric tensor $A_{\m_1\m_2\ldots\m_p}$ by the associated
$p$-form
\be
A_p\equiv
A_{\m_1\m_2\ldots\m_p}dx^{\m_1}\wedge\ldots\wedge dx^{\m_p}
\,.\label{562}\ee
Such $p$-forms transform under generalized gauge transformations:
\be
A_p\to A_p+d~\Lambda_{p-1},
\,,\label{563}\ee
where $d$ is the exterior derivative ($d^2=0$) and $\Lambda_{p-1}$ is a
$(p-1)$-form which serves as the parameter of gauge transformations.
The familiar case of (abelian) gauge fields corresponds to $p=1$.
The gauge invariant field strength is
\be
F_{p+1}=d~A_{p}
\,.\label{564}\ee
satisfying the free Maxwell equations
\be
d^{*}F_{p+1}=0
\label{5644}\ee

The natural objects, charged under a $(p+1)$-form $A_{p+1}$ are
$p$-branes.
A $p$-brane is an extended object with $p$ spatial dimensions.
Point particles correspond to $p=0$, strings to $p=1$.
The natural coupling of $A_{p+1}$ and a $p$-brane is given by
\be
\exp\left[iQ_p\int_{\rm world-volume} A_{p+1}\right]=
\exp\left[iQ_p\int A_{\m_0\ldots\m_p}dx^{\m_0}\wedge\ldots\wedge
dx^{\m_p}\right]
\,,\label{565}\ee
which generalizes the Wilson line coupling in the case of
electromagnetism.
The world-volume of $p$-brane is $p+1$-dimensional.
Note also that this is precisely the $\s$-model coupling of the usual
string
to the NS antisymmetric tensor.
The charge $Q_p$ is the usual electric charge for $p=0$ and the string
tension for $p=1$.
For the $p$-branes we will be considering, the (electric) charges will
be related to their tensions (mass per unit volume).

In analogy with electromagnetism, we can also introduce magnetic
charges.
First, we must define the analog of the magnetic field: the magnetic
(dual) form.
This is done by first dualizing the field strength and then rewriting
it as the exterior derivative of another form\footnote{This is
guaranteed by (\ref{5644}).} :
\be
d\tilde A_{D-p-3}=\tilde F_{D-p-2}=^*F_{p+2}=^*dA_{p+1}
\,,\label{566}\ee
where $D$ is the the dimension of spacetime.
Thus, the dual (magnetic) form couples to $(D-p-4)$-branes that play
the role of magnetic monopoles with ``magnetic charges" $\tilde
Q_{D-p-4}$.

There is a generalization of the Dirac quantization condition to
general
p-form charges discovered by Nepomechie and Teitelboim, \cite{nt}.
The argument parallels that of Dirac. Consider an electric p-brane
with charge $Q_p$ and a magnetic $(D-p-4)$-brane with charge $\tilde
Q_{D-p-4}$.
Normalize the forms so that the kinetic term is ${1\over 2}\int
^*F_{p+2}F_{p+2}$.
Integrating the field strength $F_{p+2}$ on a $(D-p-2)$-sphere
surrounding
the $p$-brane we obtain the total flux $\Phi=Q_p$.
We can also write
\be
\Phi=\int_{S^{D-p-2}}~^*F_{p+2}=\int_{S^{D-p-3}}~\tilde A_{D-p-3}
\,,\label{567}\ee
where we have used (\ref{566}) and we have integrated around the
``Dirac string".
When the magnetic brane circles the Dirac string it picks up a phase
$e^{i\Phi\tilde Q_{D-p-4}}$ as can be seen from (\ref{565}).
Unobservability of the string implies the Dirac-Nepomechie-Teitelboim
quantization condition
\be
\Phi \tilde Q_{D-p-4}=Q_{p}\tilde Q_{D-p-4}=2\pi N\;\;\;,\;\;\;n\in Z
\,.\label{568}\ee

\renewcommand{\theequation}{\thesection.\arabic{equation}}
\section{BPS states and bounds}
\setcounter{equation}{0}

The notion of BPS states is of capital importance in discussions of
non-perturbative duality symmetries.
Massive BPS states appear in theories with extended supersymmetry.
It just happens that sometimes supersymmetry representations are
shorter than usual. This is due to some of the supersymmetry operators
being ``null"
so that they cannot create new states.
The vanishing of some supercharges depends on the relation between the
mass of a multiplet and some central charges appearing in the
supersymmetry algebra.
These central charges depend on electric and magnetic charges of the
theory as well as expectation values of scalars (moduli).
In a sector with given charges, the BPS states are the lowest lying
states
and they saturate the so-called BPS bound which for point-like states
is of the form
\be
M\geq ~{\rm maximal~~eigenvalue~~of }\;\; Z
\,,\label{569}\ee
where $Z$ is the central charge matrix.
This is shown in appendix B where we discuss in detail the
representations of extended supersymmetry in four dimensions.

BPS states behave in very special way.

$\bullet$ At generic points in moduli space  they are absolutely
stable.
The reason is the dependence of their mass on conserved charges.
Charge and energy conservation prohibits their decay.
Consider as an example, the BPS mass formula
\be
M^2_{m,n}={|m+n\tau|^2\over \tau_2}\;\;,
\label{du2}
\ee
where $m,n$ are integer valued conserved charges, and $\tau$ is a
complex modulus. This BPS formula is relevant for N=4, SU(2) gauge
theory, in a subspace of its moduli space.
Consider a BPS state with charges $(m_0,n_0)$, at rest, decaying into N
states
with charges $(m_i,n_i)$ and masses $M_i$, $i=1,2,\cdots,N$.
Charge conservation implies, that $m_0=\sum_{i=1}^N m_i$,
$n_0=\sum_{i=1}^N n_i$.
The four-momenta of the  particles produced are $(\sqrt{M_i^2+\vec
p_i^2},\vec p_i)$ with $\sum_{i=1}^N \vec p_i=\vec 0$.
Conservation of energy implies
\be
M_{m_0,n_0}=\sum_{i=1}^N\sqrt{M_i^2+\vec p^2_i}\geq \sum_{i=1}^N
M_i\;\;.
\label{du1}\ee
Also in a given charge sector (m,n) the BPS bound implies that any mass
$M\geq M_{m,n}$ with $M_{m,n}$ given in (\ref{du2}).
Thus, from (\ref{du1}) we obtain
\be
M_{m_0,n_0}\geq \sum_{i=1}^N M_{m_i,n_i}\;\;,
\label{du3}
\ee
and the equality will hold if all  particles are BPS and are produced
at rest ($\vec p_i=\vec 0$).
Consider now the two-dimensional vectors $v_i=m_i+\tau n_i$ on the
complex $\tau$-plane, with length $||v_i||^2=|m_i+n_i\tau|^2$.
They satisfy, $v_0=\sum_{i=1}^N v_i$.
Repeated application of the triangle inequality implies
\be
||v_0||\leq \sum_{i=1}^N ||v_i||\;\;.
\label{du4}
\ee
This is incompatible with energy conservation (\ref{du3}) unless
all vectors $v_i$ are parallel. This will happen only if $\tau$ is
real.
For energy conservation it should also be a rational number.
On the other hand, due to the SL(2,$\Z$) invariance of (\ref{du2}), the
inequivalent choices for $\tau$ are in the SL(2,$\Z)$ fundamental
domain and
$\tau$ is never real there. In fact, real rational values of $\tau$ are
mapped by SL(2,$\Z)$ to $\tau_2=\infty$, and since $\tau_2$ is the
inverse of the coupling constant, this corresponds to the degenerate
case of zero coupling.
Consequently, for $\tau_2$ finite, in the fundamental domain, the BPS
states of this theory are absolutely stable. This is always true in
theories
with more than 8 conserved supercharges (corresponding to N$>2$
supersymmetry
in four dimensions).
In cases, corresponding to theories with 8 supercharges, there are
regions in the moduli space, where BPS states, stable at weak coupling,
can decay at strong coupling. However, there is always a large region
around weak coupling, where they are stable.

$\bullet$ The mass-formula of BPS states is supposed to be exact if one
uses
renormalized values for the charges and moduli.
The argument is that quantum corrections would spoil the relation of
mass and charges and if we assume unbroken SUSY at the quantum level
that would give incompatibilities  with the dimension of their
representations.
Of course this argument seems to have a loophole: a specific set of BPS
multiplets can combine into a long one. In that case, the argument
above
does not prohibit corrections.
Thus, we have to count BPS states modulo long supermultiplets.
This is precisely what helicity supertrace formulae do for us.
They are reviewed in detail in appendix B.
Even in the case of N=1 supersymmetry there is an analog of BPS states,
namely the massless states.

There are several amplitudes that in perturbation theory obtain
contributions
from BPS states only.
In the case of 8 conserved supercharges (N=2 supersymmetry in four
dimensions), all two-derivative terms as well as
$R^2$ terms are of that kind.
In the the case of 16 conserved supercharges (N=4 supersymmetry in four
dimensions) except the terms above, also the four derivative terms as
well as $R^4$, $R^2 F^2$ terms are of a similar kind.
The normalization argument of the BPS mass formula makes another
important
assumption: That as the coupling grows, there is no phase transition
during which supersymmetry is (partially) broken.

The BPS states described above can be realized as point-like soliton
solutions
of the relevant effective supergravity theory.
The BPS condition is the statement that the soliton solution leaves
part of the supersymmetry unbroken.
The unbroken generators do not change the solution, while the broken
ones
generate the supermultiplet of the soliton which is thus shorter than
the generic supermultiplet.

So far we discussed point-like BPS states. There are however BPS
versions for extended objects (BPS p-branes).
In the presence of extended objects the supersymmetry algebra can
acquire central charges that are not Lorentz scalars (as we assumed in
Appendix B).
Their general form can be obtained from group theory in which case we
deduce
that they must be antisymmetric tensors, $Z_{\m_1\ldots\m_p}$.
Such central charges have values proportional to the charges $Q_{p}$ of
 p-branes. Then, the BPS condition would relate these charges with the
energy densities (p-brane tensions) $\mu_p$ of the relevant p-branes.
Such p-branes can be viewed as extended soliton solutions of the
effective theory. The BPS condition is the statement that the soliton
solution leaves
some of the supersymmetries unbroken.

\renewcommand{\theequation}{\thesection.\arabic{equation}}
\section{Massless RR states\label{RRR}}
\setcounter{equation}{0}

We will now consider  in more detail the massless R-R states of
type-IIA,B string theory, since they have unusual properties and play a
central role
in non-perturbative duality symmetries. The reader is referred to
\cite{bac} for further reading.

I will first start by describing in detail the $\Gamma$-matrix
conventions
in flat ten-dimensional Minkowski space \cite{GSW}.

The $32\times 32$-dimensional $\Gamma$-matrices satisfy
\be
\{\Gamma^{\m},\Gamma^{\n}\}=-2\eta^{\mu\nu}\;\;\;,\;\;\;
\eta^{\m\n}=(-++\ldots +)
\,.\label{GG1}\ee
The $\Gamma$-matrix indices are raised and lowered with the flat
Minkowski
metric $\eta^{\m\n}$.
\be
\Gamma_{\m}=\eta_{\m\n}\Gamma^{\n}\;\;\;\,\;\;\;
\Gamma^{\m}=\eta^{\m\n}\Gamma_{\n}\;.
\ee
We will be in the Majorana representation where the $\Gamma$-matrices
are pure imaginary, $\Gamma^{0}$ is antisymmetric, the rest symmetric.
Also
\be
\Gamma^{0}\Gamma_{\m}^{\dagger}\Gamma^0=\Gamma_{\m}\;\;\;,\;\;\;
\Gamma^{0}\Gamma_{\m}\Gamma^0=-\Gamma_{\m}^T
\,.\label{GG2}\ee
Majorana spinors $S_{\a}$ are real: $S^{*}_{\a}=S_{\a}$.
\be
\Gamma_{11}=\Gamma_{0}\ldots\Gamma_9\;\;\;,\;\;\;(\Gamma_{11})^2=
1\;\;\;,\;\;\;
\{\Gamma_{11},\Gamma^{\m}\}=0
\,.\label{GG3}\ee
$\Gamma_{11}$ is symmetric and real.
This is the reason that in ten dimensions the Weyl condition
$\Gamma_{11}S=\pm S$
is compatible with the Majorana condition.\footnote{In a space with
signature
(p,q) the Majorana and Weyl conditions are compatible provided $|p-q|$
is a multiple of eight.}
We use the convention that for the Levi-Civita tensor, $\e^{01\ldots
9}=1$.
We will define the antisymmetrized products of $\Gamma$-matrices
\be
\Gamma^{\m_1\ldots\m_k}={1\over k!}\Gamma^{[\m_1}\ldots\Gamma^{\m_k]}=
{1\over k!}\left(\Gamma^{\m_1}\ldots\Gamma^{\m_k}\pm{\rm
permutations}\right)
\,.\label{GG4}\ee

We can derive by straightforward computation the following
identities among $\Gamma$-matrices:
\bea
\Gamma_{11} \Gamma^{\mu_1...\mu_k} &=&
{(-1)^{[{k\over 2}]} \over (10-k)!}
 \epsilon^{\mu_1...\mu_{10}} \Gamma_{\mu_{k+1}...\mu_{10}}\;,
\\
\Gamma^{\mu_1...\mu_k} \Gamma_{11} &=&
 {(-1)^{[{k+1\over 2}]}\over (10-k)!}
 \epsilon^{\mu_1...\mu_{10}} \Gamma_{\mu_{k+1}...\mu_{10}}
\,,\label{GG5}\eea
with $[x]$ denoting the integer part of $x$.
\be
\Gamma^\mu \Gamma^{\nu_1...\nu_k} =  \Gamma^{\mu\nu_1...\nu_k}
-{1\over (k-1)!} \eta^{\mu [\nu_1} \Gamma^{\nu_2...\nu_k]}
\,,\label{GG6}\ee
\be
 \Gamma^{\nu_1...\nu_k} \Gamma^\mu =  \Gamma^{ \nu_1...\nu_k\mu}
-{1\over (k-1)!} \eta^{\mu [\nu_k} \Gamma^{\nu_1...\nu_{k-1}]}
\,,\label{GG7}\ee
with square brackets denoting the  alternating sum
over all permutations of the enclosed indices.
The invariant Lorentz scalar product of two spinors $\chi,\phi$ is
$\chi^{*}
_{\a}(\Gamma^{0})_{\a\b}\phi_{\b}$.

Now consider the ground-states of the Ramond-Ramond sector.
On the left, we have a Majorana spinor $S_{\a}$ satisfying $\Gamma_{11}
S=S$ by convention.
On the right we have another Majorana spinor $\tilde S_{\a}$ satisfying
$
\Gamma_{11}\tilde S=\xi \tilde S$ where $\xi=1$ for the type-IIB string
and $\xi=-1$ for the type-IIA string.
The total ground-state is the product of the two.
To represent it, it  is convenient to define the following bispinor
field
\be
F_{\a\b}=S_{\a}(i\Gamma^{0})_{\b\g}\tilde S_{\g}
\,.\label{GG8}\ee
With this definition, $F_{\a\b}$ is real and the trace
$F_{\a\b}\delta^{\a\b}$ is Lorentz invariant.
The chirality conditions on the spinor translate into
\be
\Gamma_{11}F=F\;\;\;,\;\;\;F\Gamma_{11}=-\xi F
\,,\label{gg9}\ee
where we have used that $\Gamma_{11}$ is symmetric and anticommutes
with $\Gamma^0$.

We can now expand the bispinor $F$ into the complete set of
antisymmetrized $\Gamma$'s
\be
F_{\a\b}=\sum_{k=0}^{10}{i^k\over k!}F_{\m_1\ldots\m_k}
(\Gamma^{\m_1\ldots\m_k})_{\a\b}
\,,\label{gg10}\ee
where the $k=0$ term is proportional to the unit matrix and the tensors
$F_{\m_1\ldots \m_k}$ are real.

We can now translate the first of the chirality conditions in
(\ref{gg9})
using (\ref{GG5}) to obtain the following equation:
\be
F^{\m_1\ldots\m_k}={(-1)^{\left[{k+1\over 2}\right]}\over (10-k)!}\e
^{\m_1\ldots\m_{10}}F_{\m_{k+1}\ldots\m_{10}}
\,.\label{gg11}\ee
The second chirality condition implies
\be
F^{\m_1\ldots\m_k}=\xi{(-1)^{\left[{k\over 2}\right]+1}\over (10-k)!}\e
^{\m_1\ldots\m_{10}}F_{\m_{k+1}\ldots\m_{10}}
\,.\label{gg12}\ee
Compatibility between (\ref{gg11}) and (\ref{gg12}) implies that
type-IIB
theory ($\xi=1$) contains tensors of odd rank (the independent ones
being
k=1,3 and k=5 satisfying a self-duality condition) and type-IIA theory
($\xi=-1)$
contains tensors of even rank (the independent ones having k=0,2,4).
The number of independent tensor components adds up in both cases to
$16\times 16=256$.

The mass-shell
conditions $G_0=\bar G_0=0$ imply that the bispinor field (~\ref{GG1})
obeys two
 massless Dirac equations coming from $G_0$ and $\bar G_0$:
\be
 (p_\mu \Gamma^\mu) F = F (p_\mu \Gamma^\mu) = 0 \
\,.\label{G6}\ee
To convert these to equations for the tensors we use the gamma
identities (\ref{GG6},\ref{GG7}).
After some
straightforward algebra one finds
\be
 p^{[\mu} F^{\nu_1...\nu_k]} = p_\mu F^{\mu \nu_2...\nu_k} =0
\,,\label{G9}\ee
which are the Bianchi identity and free massless equation for
an antisymmetric tensor field strength. We may write these in
economic form as
\be
 d F = d \ ^*F = 0
\,.\label{G10}\ee
Solving the Bianchi identity locally
allows us to express the  $k$-index field strength as the
exterior derivative of a $(k-1)$-form potential
\be
F_{\mu_1...\mu_k} =
{1\over (k-1)!} \partial_{[\mu_1} C_{\mu_2...\mu_k]}
\,,\label{G11}\ee
or in short-hand notation
\be
 F_{(k)} = d C_{(k-1)}
\,.\label{G12}\ee
Consequently, the type-IIA theory has a vector ($C^\mu$) and a
three-index
tensor potential ($C^{\mu\nu\rho}$) , in addition to a constant
non-propagating zero-form field strength ($F$), while the
type-IIB theory has a zero-form ($C$), a two-form ($C^{\mu\nu}$)
and a four-form potential ($C^{\mu\nu\rho\sigma}$), the latter
 with self-dual field strength. The number of physical transverse
degrees of freedom adds up in both cases to $64=8\times 8$.

It is not difficult to see that in the perturbative string
spectrum there are no states charged under the RR forms.
First, couplings of the form $\langle s|R\overline{R}|s\rangle$ are not
allowed
by the separately conserved left and right fermion numbers.
Second, the RR vertex operators contain the field strengths rather than
the potentials and equations of motion and Bianchi identities enter on
an equal footing. If there were electric states in perturbation theory
we would also
have magnetic states.

RR forms have another peculiarity.
There are various ways to deduce that their couplings to the dilaton
are exotic.
The dilaton dependence of a $F^{2m}$ term at the k-th order of
perturbation theory is $e^{(k-1)\Phi}e^{m\Phi}$ instead of the usual
$e^{(k-1)\Phi}$ term for NS-NS fields.
For example, at tree-level, the quadratic terms are dilaton
independent.

\renewcommand{\theequation}{\thesection.\arabic{equation}}
\section{Heterotic/Type-I duality in ten dimensions.\label{hetI}}
\setcounter{equation}{0}

We will start our discussion by describing heterotic/type-I duality
in ten dimensions.
It can be shown \cite{sen} that heterotic/type-I duality, along with
T-duality can reproduce all known string dualities.

Consider first the O(32) heterotic string theory.
At tree-level (sphere) and up to two-derivative terms, the (bosonic)
effective
action in the $\s$-model frame is
\be
S^{\rm het}=\int d^{10}x\sqrt{G}e^{-\Phi}\left[
R+(\nabla\Phi)^2-{1\over 12}\hat H^2-{1\over 4}F^2\right]
\,.\label{570}\ee

On the other hand for the O(32) type I string the leading order
two-derivative effective action is
\be
S^{I}=\int d^{10}x\sqrt{G}\left[e^{-\Phi}\left(
R+(\nabla\Phi)^2\right)-{1\over 4}e^{-\Phi/2}F^2-{1\over 12}\hat
H^2\right]
\,.\label{571}\ee
The different dilaton dependence here comes as follows: The Einstein
and dilaton terms come from the closed sector on the sphere ($\chi=2$).
The gauge kinetic terms come from the disk ($\chi=1$). Since the
antisymmetric tensor comes from the RR sector of
the closed superstring it does not have any dilaton
dependence on the sphere.

We will now bring both actions to the Einstein frame,
$G_{\m\n}=e^{\Phi/4}g_{\m\n}$:
\be
S^{\rm het}_E=\int d^{10}x\sqrt{g}\left[
R-{1\over 8}(\nabla\Phi)^2-{1\over 4}e^{-\Phi/4}F^2-{1\over
12}e^{-\Phi/2}\hat H^2\right]
\,,\label{572}\ee
\be
S^{I}_E=\int d^{10}x\sqrt{g}\left[
R-{1\over 8}(\nabla\Phi)^2-{1\over 4}e^{\Phi/4}F^2-{1\over
12}e^{\Phi/2}\hat H^2\right]
\,.\label{573}\ee

We observe that the two actions are related by $\Phi\to -\Phi$ while
keeping the other fields invariant. This seems to suggest that the weak
coupling of one
is the strong coupling of the other and vice versa.
Of course, the fact that the two actions are related by a field
redefinition
is not a surprise. It is known that N=1 ten-dimensional supergravity is
completely fixed once the gauge group is chosen.
It is interesting though that the field redefinition here is just an
inversion of the ten-dimensional coupling.
Moreover, the two theories have perturbative expansions that are very
different.

We would like to go further and check if there are non-trivial checks
of
what is suggested by the classical N=1 supergravity.
However, once we compactify one direction on a circle of radius $R$ we
seem
to have a problem.
In the heterotic case, we have a spectrum that depends both on momenta
$m$ in the ninth direction as well as on windings $n$.
The winding number is the charge that couples to the string
antisymmetric tensor. In particular, it is the electric charge of the
gauge boson obtained from
$B_{9\m}$.
On the other hand, in type I theory, as we have shown earlier, we have
momenta $m$ but no windings.
One way to see this, is that the open string Neumann boundary
conditions forbid the string to wind around the circle.
Another way is by noting that the NS-NS antisymmetric tensor that could
couple to windings has been projected out by our orientifold
projection.

However, we do have the RR antisymmetric tensor, but as we argue in
section \ref{RRR},
no perturbative states are charged under it.
There may be however non-perturbative states that are charged under
this antisymmetric tensor.
According to our general discussion in section \ref{antisym} this
antisymmetric tensor would couple naturally to a string but this is
certainly not
the perturbative string.
How can we construct this non-perturbative string?

An obvious guess is that this is a solitonic string excitation of the
low energy type-I effective action. Indeed, such a  solitonic solution
was constructed \cite{dab} and shown to have the correct zero mode
structure.

We can give a more complete description of this non-perturbative
string.
The hint is given from $T$-duality on the heterotic side, that
interchanges windings and momenta.
When it acts on derivatives of $X$ it interchanges
$\pd_{\s}X\leftrightarrow \pd_{\tau}X$.
Consequently, Neumann boundary conditions are interchanged with
Dirichlet ones.
To construct such a non-perturbative string we would have to use also
Dirichlet boundary conditions.
Such boundary conditions imply that the open string boundary in fixed
in spacetime. In terms of waves traveling on the string, it implies
that a
wave arriving at the boundary is reflected with a minus sign.
The interpretation of fixing the open string boundary in some
(submanifold)
of spacetime has the following interpretation: There is a solitonic
(extended)
object there whose fluctuations are described by open strings attached
to it.
Such objects are known today as D-branes.

Thus, we would like to describe our non-perturbative string as a
D1-brane.
We will localize it to the hyperplane $X^2=X^3=\ldots=X^9=0$. Its
world-sheet extends in the $X^0,X^1$ directions.
Such an object is schematically shown in Fig. \ref{f21}.
Its fluctuations can be described by two kinds of open strings:

\begin{figure}
\begin{center}
\leavevmode
\epsfbox{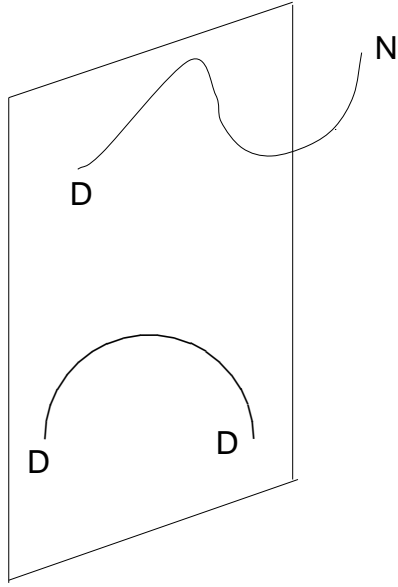}
\caption[]{Open string fluctuations of a D1-brane}
\label{f21}
\end{center}\end{figure}

$\bullet$ DD strings which have D-boundary conditions on both
end-points and are forced to move on the D1-brane.

$\bullet$ DN strings which have a D-boundary condition on one end,
which is stuck on the D1-brane, and N-boundary conditions on the other
end, which is free.

As we will see, this solitonic configuration breaks half of N=2
spacetime supersymmetry possible in ten dimensions.
It also breaks SO(9,1)$\to$SO(8)$\times$SO(1,1). Moreover, we can put
it anywhere
in the transverse eight-dimensional space, so we expect 8 bosonic
zero-modes
around it associated with the broken translational symmetry.
We will try to understand in more detail the modes describing the
world-sheet theory of the D1 string.
We can obtain them by looking at the massless spectrum of the open
string fluctuations around it.

Start with the DD strings.
Here $X^I,\psi^I,\bar\psi^I$, $I=2,\ldots,9$ have DD boundary
conditions while $X^{\m},\psi^{\m},\bar\psi^{\m}$, $\m=0,1$ have NN
boundary conditions.

For the world-sheet fermions NN boundary conditions imply
\be
{\rm NN~~NS~~sector}~~~~~~~~~\left.\psi+\bar\psi\right|_{\s=0}=\left.
\psi-\bar\psi\right|_{\s=\pi}=0
\,,\label{576}\ee
\be
{\rm NN~~R~~sector}~~~~~~~~~\left.\psi-\bar\psi\right|_{\s=0}=
\left.\psi-\bar\psi\right|_{\s=\pi}=0
\,,\label{577}\ee
The DD boundary condition is essentially the same with $\bar\psi\to-
\bar\psi$
\be
{\rm DD~~NS~~sector}~~~~~~~~~\left.\psi-\bar\psi\right|_{\s=0}=\left.
\psi+\bar\psi\right|_{\s=\pi}=0
\,,\label{580}\ee
\be
{\rm DD~~R~~sector}~~~~~~~~~\left.\psi+\bar\psi\right|_{\s=0}=
\left.\psi+\bar\psi\right|_{\s=\pi}=0
\,,\label{581}\ee
and a certain action on the Ramond ground-state that we will describe
below.

\vskip .4cm
\noindent\hrulefill
\vskip .4cm

{\large\bf Exercise} Show that we have the following mode expansions
\be
X^I(\s,\t)=x^I+w^I\s+2\sum_{n\not= 0}{a^I_{n}\over n}e^{in\tau}\sin
(n\s)
\,,\label{574}\ee
\be
X^{\m}(\s,\t)=x^{\m}+p^{\m}\t-2i\sum_{n\not= 0}{a^{\m}_{n}\over
n}e^{in\tau}\cos (n\s)
\,.\label{575}\ee
In the NS sector
\be
\psi^I(\s,\t)=\sum_{n\in Z}b^I_{n+1/2}e^{i(n+1/2)(\s+\t)}\;\;\;,\;\;\;
\psi^{\m}(\s,\t)=\sum_{n\in Z}b^{\m}_{n+1/2}e^{i(n+1/2)(\s+\t)}
\,,\label{578}\ee
while in the R sector
\be
\psi^I(\s,\t)=\sum_{n\in Z}b^I_{n}e^{in(\s+\t)}\;\;\;,\;\;\;
\psi^{\m}(\s,\t)=\sum_{n\in Z}b^{\m}_{n}e^{in(\s+\t)}
\,.\label{579}\ee
Also
\be
\bar b^I_{n+1/2}=b^I_{n+1/2}\;\;\;,\;\;\;\bar b^I_{n}=-b^I_{n}
\,,\label{584}\ee
\be
\bar b^{\m}_{n+1/2}=-b^{\m}_{n+1/2}\;\;\;,\;\;\;\bar b^{\m}_{n}=
b^{\m}_{n}
\,.\label{585}\ee

\vskip .4cm
\noindent\hrulefill
\vskip .4cm

The $x^I$ in (\ref{574}) are the position of the D-string in transverse
space.
There is no momentum in (\ref{574}) which means that the state
wavefunctions would depend only on the $X^{0,1}$ coordinates, since
there is a continuous momentum in (\ref{575}). Thus, the states of this
theory ``live" on the world-sheet of the D1-string.
The usual bosonic massless spectrum would consist of a vector
$A_{\m}(x^0,x^1)$ corresponding to the state
$\psi^{\m}_{-1/2}|0\rangle$ and eight bosons
$\phi^I(x^0,X^1)$ corresponding to the states
$\psi^{I}_{-1/2}|0\rangle$\footnote{The GSO projection is always
present.}.
We will  now consider the action of the orientation reversal $\Omega$:
$\s\to -\s$, $\psi\leftrightarrow \bar \psi$.
Using (\ref{576}-\ref{581})
\be
\Omega~b^{\m}_{-1/2}|0\rangle= \bar b^{\m}_{-1/2}|0\rangle=-
b^{\m}_{-1/2}|0\rangle
\,,\label{582}\ee
\be
\Omega~b^{I}_{-1/2}|0\rangle= \bar b^{I}_{-1/2}|0\rangle=
b^{I}_{-1/2}|0\rangle
\,.\label{583}\ee
The vector is projected out, while the eight bosons survive the
projection.

We will now analyze the Ramond sector where fermionic degrees of
freedom would come from.
The massless ground-state $|R\rangle$ is an SO(9,1) spinor satisfying
the usual
GSO projection
\be
\Gamma^{11}|R\rangle=|R\rangle
\,.\label{586}\ee
Consider now the $\Omega$ projection on that spinor.
In the usual $NN$ case $\Omega$ can be taken to commute with $(-1)^F$
and acts on the spinor ground-state as -1.
In the DD case the action of $\Omega$ on the transverse DD fermionic
coordinates is reversed compared to the NN case.
On the spinor this action is
\be
\Omega |R\rangle =-\Gamma^2\ldots\Gamma^9|R\rangle =|R\rangle
\,.\label{587}\ee
{}From (\ref{586},\ref{587}) we also obtain
\be
\Gamma^0\Gamma^1|R\rangle=-|R\rangle
\,.\label{588}\ee
If we decompose the spinor under SO(8)$\times$SO(1,1) the surviving
piece transforms as $8_-$ where $-$ refers to the SO(1,1) chirality
(\ref{588}).
As for the bosons, these fermions are functions of $X^{0,1}$ only.

To recapitulate, in the DD sector we have found the following massless
fluctuations moving on the world-sheet of the D1-string: 8 bosons and 8
chirality minus fermions.

Consider now the DN fluctuations.
In this case Chan-Patton factors are allowed in the free string end,
and the usual tadpole cancellation argument implies there are 32 of
them.
In this case, the boundary conditions for the transverse bosons and
fermions become
\be
\left.\pd_{\t}X^I\right|_{\s=0}=0\;\;\;,\;\;\;
\left.\pd_{\s}X^I\right|_{\s=\pi}=0
\,,\label{589}\ee
\be
{\rm DN~~NS~~sector}~~~~~~~~~\left.\psi+\bar\psi\right|_{\s=0}=\left.
\psi+\bar\psi\right|_{\s=\pi}=0
\,,\label{590}\ee
\be
{\rm DN~~R~~sector}~~~~~~~~~\left.\psi-\bar\psi\right|_{\s=0}=
\left.\psi+\bar\psi\right|_{\s=\pi}=0
\,,\label{591}\ee
while they are NN in the longitudinal directions.

We observe that here, the bosonic oscillators are half-integrally moded
as in the twisted sector of $Z_2$ orbifolds. Thus, the ground-state
conformal weight is 8/16=1/2.
Also the moding for the fermions has been reversed between the NS and R
sectors.
In the NS sector the fermionic ground-state is also a spinor with
ground state conformal weight 1/2. The total ground-state has
conformal weight
one and only massive excitations are obtained in this sector.

In the R sector there are massless states coming from the bosonic
ground-state combined with the O(1,1) spinor ground-state from the
longitudinal
Ramond fermions.
The usual GSO projection here is $\Gamma^0\Gamma^1=1$.
Thus, the massless modes in the DN sector are 32 chirality plus
fermions.

In total, the world-sheet theory of the D-string contains exactly what
we would
expect from the heterotic string in the physical gauge!
This is a non-trivial argument in favor of heterotic-type I duality.

\vskip .4cm
\noindent\hrulefill
\vskip .4cm

{\large\bf Exercise}. We have considered so far a D1-brane in Type I
theory.
Consider the general case of Dp-branes along similar lines.
Show that non-trivial configurations exist (compatible with GSO and
$\Omega$ projections) preserving half of the supersymmetry,  for
p=1,5,9.
The case p=9 corresponds to the usual open strings moving in 10-d
space.

\vskip .4cm
\noindent\hrulefill
\vskip .4cm

The RR two-form couples to a one-brane (electric) and a five-brane
(magnetic). As we saw above, both can be constructed as D-branes.

We will describe now in some more detail the D5-brane, since it
involves
some novel features.
To construct a five-brane, we will have to impose Dirichlet boundary
conditions
in four transverse directions.
We will again have DD and NN sectors as in the D1 case.
The massless fluctuations will have continuous momentum in the six
longitudinal
directions, and will describe fields living on the six-dimensional
world-volume of the five-brane.
Since we are breaking half of the original supersymmetry, we expect
that the
world-volume theory will have N=1 six-dimensional supersymmetry, and
the massless fluctuations will form multiplets of this supersymmetry.
The relevant multiplets are the vector multiplet, containing a vector
and a
gaugino, as well as the hypermultiplet, containing four real scalars
and
a fermion.
Supersymmetry implies that the manifold of the hypermultiplet scalars
is a hyper-K\"ahler manifold. When the hypermultiplets are charged
under the
gauge group, the gauge transformations are isometries of the
hyper-K\"ahler
manifold, of a special type: they are compatible with the
hyper-K\"ahler structure.

It will be important for our latter purposes to describe the Higgs
effect in this case.
When a gauge theory is in the Higgs phase, the gauge bosons become
massive
by combining with some of the massless Higgs modes.
The low-energy theory (for energies well below the gauge boson mass)
is described by the scalars that have not been  devoured by the gauge
bosons.
In our case, each (six-dimensional) gauge boson that becomes massive,
will
eat-up four scalars (a hypermultiplet). The left-over low-energy theory
of the scalars will be described by a smaller hyper-K\"ahler manifold
(since supersymmetry is not broken during the Higgs phase transition).
This manifold is constructed by a mathematical procedure known as the
hyper-K\"ahler quotient.
The procedure "factors out" the isometries of a hyper-K\"ahler manifold
to produce a lower dimensional manifold which is still hyper-K\"ahler.
Thus, the  hyper-K\"ahler quotient construction is describing the
ordinary
Higgs effect in six-dimensional N=1 gauge theory.

The D5-brane we are about to construct, is mapped via heterotic/type-I
duality
to the NS5-brane of the heterotic theory. The NS5-brane, has been
constructed
\cite{chs} as a soliton of the effective low-energy heterotic action.
The non-trivial fields, in the transverse space, are essentially
configurations
of axion-dilaton instantons, together with four-dimensional instantons
embedded in the O(32) gauge group. Such instantons have a size that
determines
the ``thickness" of the NS5-brane.
The massless fluctuations are essentially the moduli of the instantons.
There is a mathematical construction of this moduli space, as a
hyper-K\"ahler
quotient. This leads us to suspect \cite{si} that the interpretation of
this construction  is a Higgs
effect in the six-dimensional world volume theory.
In particular, the mathematical construction implies that for N
coincident
NS5 branes, the hyper-K\"ahler quotient construction implies that an
Sp(N) gauge group is completely Higgsed. For a single five-brane, the
gauge group
is $\rm Sp(1)\sim SU(2)$.
Indeed, if the size of the instanton is not zero, the massless
fluctuations
of the NS5-brane form hypermultiplets only.
When, the size becomes zero, the moduli space has a singularity, which
can be interpreted as the restoration of the gauge symmetry: at this
point
the gauge bosons become massless again.
All of this indicates that the world-volume theory of a single
five-brane
should contain an SU(2) gauge group, while in the case of N five-branes
the gauge group is enhanced to Sp(N), \cite{si}.

We will return now in our description of the massless fluctuations of
the D5-brane.
The situation parallels the D1 case that we have described in detail.
In particular, from the DN sectors we will obtain hypermultiplets only.
{}From the DD sector we can in principle obtain massless vectors.
However, as we have seen above, the unique vector that can appear is
projected
out by the orientifold projection.
To remedy this situation we are forced to introduce a Chan-Patton
factor
for the Dirichlet end-points of the open string  fluctuations.
For a single D5-brane, this factor takes two values, $i=1,2$.
Thus, the massless bosonic states in the DD sector are of the form,
\be
b_{-1/2}^{\mu}|p;i,j\rangle\;\;\;\;\;\;,\;\;\;\;\;\;
b_{-1/2}^{I}|p;i,j\rangle\;.
\ee
We have also seen, that the orientifold projection $\Omega$ changes the
sign of $b_{-1/2}^{\mu}$ and leaves $b_{-1/2}^{I}$ invariant.
The action of $\Omega$ on the ground state is $\Omega
|p;i,j\rangle=\e|p;j,i\rangle$. It interchanges the Chan-Patton factors
and can have a sign $\e=\pm 1$.
The number of vectors that survive the $\Omega$ projection depends on
this sign.
For $\e=1$, only one vector survives and the gauge group is O(2).
If $\e=-1$, three vectors survive and the gauge group is $\rm Sp(1)\sim
SU(2)$.
Taking into account our previous discussion, we must take $\e=-1$.
Thus, we have an Sp(1) vector multiplet.
The scalar states on the other hand will be forced to be
antisymmetrized in
the Chan-Patton indices. This will provide  a single hypermultiplet,
whose four scalars
describe the position of the D5-brane in the four-dimensional
transverse space.
Finally, the DN sector, has an $i=1,2$ Chan-Patton factor on the D-end
and an $\a=1,2\cdots 32$ factor on the N-end. Consequently, we will
obtain
a hypermultiplet transforming as $\bf (2,32)$ under $\rm Sp(1)\times
O(32)$
where Sp(1) is the world-volume gauge group and O(32) is the original
(spacetime) gauge group of the type-I theory.

In order  to describe N parallel coinciding D5-branes, the only
difference is that the Dirichlet Chan-Patton factor now takes 2N
values.
Going through the same procedure as above we find in the DD sector,
Sp(N) vector multiplets,
and hypermultiplets transforming as a singlet (the center of mass
position coordinates) as well as the traceless symmetric tensor
representation of Sp(N)
of dimension $2N^2-N-1$.
In the DN sector we find a hypermultiplet transforming as $\bf (2N,32)$
under
$\rm Sp(N)\times O(32)$.

There are further checks of heterotic/type-I duality in ten dimensions.
BPS saturated terms in the effective action match appropriately between
the two theories \cite{ts}.
You can find  a more detailed exposition of similar matters in
\cite{P2}.

The comparison becomes more involved and non-trivial  upon toroidal
compactification.
First, the spectrum of BPS states is richer and different in
perturbation theory in the two theories. Second, by adjusting moduli
both theories can be compared in the weak coupling limit.
The terms in the effective action that can be easiest compared are the
$F^4$,
$F^2R^2$ and $R^4$ terms. These are BPS saturated and anomaly related.
In the heterotic string, they obtain perturbative corrections at
one-loop only.
Also, their non-perturbative corrections are due to instantons that
preserve half of the supersymmetry. Corrections due to generic
instantons, that break
more than 1/2 supersymmetry, vanish due to zero modes.
In the heterotic string the only relevant non-perturbative
configuration
is the NS5-brane. Taking its world-volume to be Euclidean and wrapping
it
supersymmetrically around a compact manifold (so that the classical
action is finite), it provides the relevant instanton configurations.
Since we need at least a six-dimensional compact manifold to wrap it,
we can immediately deduce that the BPS saturated terms do not have
non-perturbative corrections for toroidal compactifications with more
than four non-compact directions.
Thus, for $D>4$ the full heterotic result is tree-level and one-loop.

In the type-I string the situation is slightly different.
Here we have both the D1-brane and the D5-brane, that can provide
instanton configurations. Again, the D5-brane will contribute in four
dimensions.
However, the D1-brane has a two-dimensional world-sheet and can
contribute already in eight dimensions.
We conclude that in nine-dimensions, the two theories can be compared
in perturbation theory.
This has been done in \cite{bk1}. They do agree at one-loop. On the
type-I
side however, duality implies also contact contributions for
the factorizable terms $(tr R^2)^2$, $tr F^2tr R^2$ and $(tr F^2)^2$
coming
from surfaces with Euler number $\chi=-1,-2$.

In eight dimensions, the perturbative heterotic result, is mapped via
duality to perturbative as well as non-perturbative type I
contributions coming from the D1-instanton.
These have been computed and duality has been verified \cite{bk2}.

\renewcommand{\theequation}{\thesection.\arabic{equation}}
\section{Type-IIA versus M-theory.}
\setcounter{equation}{0}

We have mentioned in an earlier section, that the effective type-IIA
supergravity is the dimensional reduction of eleven-dimensional, N=1
supergravity.
We will see here that this is not just an accident \cite{ht,va}.

We will first review the spectrum of forms in type-IIA theory in ten
dimensions.

$\bullet$ NS-NS two-form B. Couples to a string (electrically) and a
five-brane (magnetically). The string is the perturbative type-IIA
string.

$\bullet$ RR U(1) gauge field A$_{\mu}$. Can couple electrically to
particles
(zero-branes) and magnetically to six-branes. Since it comes from the
RR sector
no perturbative state is charged under it.

$\bullet$ RR three-form C$_{\m\n\rho}$. Can couple electrically to
membranes (p=2) and magnetically to four-branes.

$\bullet$ There is also the non-propagating zero-form field strength
and
ten-form field strength that would couple to eight-branes (see section
\ref{RRR}).

The lowest-order type-IIA Lagrangian is
\be
\tilde S^{IIA}={1\over 2\kappa^2_{10}}\left[\int
d^{10}x\sqrt{g}e^{-\Phi}\left[\left(R+(\nabla\Phi)^2-{1\over
12}H^2\right)-{1\over 2\cdot 4!}\hat G^2 -{1\over 4}F^2\right]+{1\over
(48)^2}\int B\wedge G\wedge G\right].\label{596}\ee
We are in the string frame.
Note that the RR kinetic terms do not couple to the dilaton as argued
already in section \ref{RRR}.

In the type-IIA supersymmetry algebra there is a central charge
proportional to the U(1) charge of the gauge field A
\be
\{Q^1_{\a},Q^2_{\dot \a}\}=\d_{\a\dot\a} W
\,.\label{595}\ee
This can be understood, since this supersymmetry algebra is coming from
D=11
where instead of $W$ there is the momentum operator of the eleventh
dimension. Since the U(1) gauge field is the $G_{11,\m}$ component of
the metric, the momentum operator becomes the U(1) charge in the
type-IIA theory.
There is an associated BPS bound
\be
M\geq {c_0\over \lambda}|W|
\,,\label{597}\ee
where $\lambda=e^{\Phi/2}$ is the ten-dimensional string coupling and
$c_0$ some constant.
States that satisfy this equality are BPS saturated and form smaller
supermultiplets.
As mentioned above all perturbative string states have $W=0$.
However, there is a soliton solution (black hole) of type-IIA
supergravity with the required properties.
In fact, the BPS saturation implies that it is an extremal black hole.
We would expect that quantization of this solution would provide a
(non-perturbative) particle state.
Moreover, it is reasonable to expect that the U(1) charge is quantized
in some units.
Then the spectrum of these BPS states looks like
\be
M ={c\over \lambda}|n|\;\;\;,\;\;\;n\in Z
\,.\label{598}\ee
At weak coupling these states are very heavy (but not as heavy as
standard solitons whose masses scale with the coupling as $1/\l^2$).
However, being BPS states, their mass can be reliably followed at
strong coupling where they become light, piling up at zero mass as the
coupling becomes infinite.
This is precisely the behavior of Kaluza-Klein (momentum) modes as a
function of the radius. Since also the effective type-IIA field theory
is a dimensional reduction of the eleven-dimensional supergravity with
$G_{11,11}$ becoming the string coupling,
we can take this seriously \cite{va} and claim that as $\l\to \infty$
type-IIA theory becomes some
eleven-dimensional theory whose low energy limit is eleven-dimensional
supergravity.
We can calculate the relation between the radius of the eleventh
dimension
and the string coupling.

The N=1 eleven-dimensional supergravity action is
\be
L^{D=11}={1\over 2\kappa^2}\left[R-{1\over 2\cdot
4!}G_4^2\right]-i\bar\psi_{\mu}\Gamma^{\mu\nu\rho}
\tilde\nabla_{\nu}\psi_{\rho}+{1\over 2\kappa(144)^2}
G_4\wedge G_4\wedge \hat C+
\label{300}\ee
$$
+{1\over 192}\left[\bar
\psi_{\mu}\Gamma^{\mu\nu\rho\s\tau\upsilon}\psi_{\upsilon}
+12\bar\psi^{\nu}\Gamma^{\rho\s}\psi^{\tau}\right]
(G+\hat G)_{\nu\rho\s\tau}\,,
$$
where $\tilde \nabla$ is defined with respect to the connection
$(\omega+\tilde\omega)/2$, $\omega$ is the spin connection and
\be
\tilde \omega_{\mu,ab}=\omega_{\mu,ab}+{i\kappa^2\over
4}\left[-\bar\psi^{\nu}\Gamma_{\nu\mu
ab\rho}\psi^{\rho}+2(\bar\psi_{\mu}
\Gamma_b\psi_{a}-\bar\psi_{\mu}
\Gamma_a\psi_{b}+\bar\psi_{b}
\Gamma_{\mu}\psi_{a})\right]
\label{301}\ee
is its supercovariantization.
Finally, $G_4$ is the field strength of $\hat C$,
\be
G_{\m\n\rho\s}=\pd_{\mu}\hat C_{\n\rho\s}-\pd_{\nu}\hat C_{\rho\s\mu}
+\pd_{\rho}\hat C_{\s\m\n}-\pd_{\s}\hat C_{\m\n\rho}
\ee
and $\tilde G_4$ is its supercovariantization
\be
\tilde G_{\m\n\rho\s}=G_{\m\n\rho\s}-6\kappa^2\bar
\psi_{[\m}\Gamma_{\n\rho}
\psi_{\s]}\;.
\ee
We will dimensionally reduce to D=10.
\be
G_{\mu\nu}=\left(\matrix{g_{\mu\nu}+
e^{2\s}A_{\mu}A_{\nu}&e^{2\s}A_{\mu}\cr
e^{2\s}A_{\mu}&e^{2\s}\cr}\right)
\,.\label{302}\ee
to be $R=e^{\s}$.
The three-form $\hat C$ gives rise to a three-form and a two-form in
ten dimensions
\be
C_{\mu\nu\rho}=\hat C_{\mu\nu\rho}-\left(\hat
C_{\nu\rho,11}A_{\mu}+{\rm cyclic}\right)\;\;,\;\;
B_{\mu\nu}=\hat C_{\mu\nu,11}\;.
\ee

The ten-dimensional action can be directly obtained from the
eleven-dimensional one using the formulae of Appendix A.
For the bosonic part we obtain,
\be
S^{IIA}={1\over 2\kappa^2}\int d^{10}x\sqrt{g}e^{\s}\left[R-{1\over
2\cdot 4!}\hat G^2-{1\over 2\cdot 3!}e^{-2\s}H^2 -{1\over
4}e^{2\s}F^2\right]+\label{6065}\ee
$$
+{1\over 2\kappa (48)^2}\int B\wedge G\wedge G\;,
$$
where
\be
F_{\mu\nu}=\pd_{\m}A_{\n}-\pd_{\n}A_{\m}\;\;\;,\;\;\;H_{\mu\nu\rho}=
\pd_{\mu}B_{\nu\rho}+{\rm cyclic}\;,
\ee
\be
\hat G_{\m\n\rho\s}=G_{\mu\nu\rho\sigma}+(F_{\mu\nu}B_{\rho\sigma}+{\rm
5~~ permutations})\;.
\ee
This is the type-IIA effective action in the Einstein frame.
We can go to the string frame by $g_{\m\n}\to e^{-\s}g_{\m\n}$. The
ten-dimensional dilaton is   $\Phi=3\s$.
The action is
\be
\tilde S_{10}={1\over 2\kappa^2}\int
d^{10}x\sqrt{g}e^{-\Phi}\left[\left(R+(\nabla\Phi)^2-{1\over
12}H^2\right)-{1\over 2\cdot 4!}\hat G^2 -{1\over 4}F^2\right]+{1\over
2\kappa^2 (48)^2}\int B\wedge G\wedge G
\,.\label{6055}\ee
Note that the kinetic terms of the RR fields $A_{\mu}$ and
$C_{\m\n\rho}$
do not have dilaton dependence at the tree level, as advocated in
section
\ref{RRR}.

The radius of the eleventh dimension is given by $R=e^{\s}$.
Thus,
\be
R=\l^{2/3}
\,.\label{604}\ee

At strong type-IIA coupling,  $R\to\infty$ and the theory
decompactifies to eleven dimensions, while in the perturbative regime
the radius is small.

The eleven-dimensional theory (which has been named M-theory) contains
the three-form which can couple to a membrane and a five-brane.
Upon toroidal compactification to ten dimensions ,
the membrane, wrapped around the circle, becomes the perturbative
type-IIA string that couples to $B_{\m\n}$.
When it is not winding around the circle, then it is the type-IIA
membrane coupling to the type-IIA three-form.
The M-theory five-brane descends to the type-IIA five-brane or, wound
around the circle, to the type-IIA four-brane.

\renewcommand{\theequation}{\thesection.\arabic{equation}}
\section{M-theory and the E$_8\times$E$_8$ heterotic string.}
\setcounter{equation}{0}

M-theory has $Z_2$ symmetry under which the three-form changes sign.
We might consider an orbifold of M-theory compactified on a circle of
radius R, where the orbifolding symmetry
is $x^{11}\to -x^{11}$ as well as the $Z_2$ symmetry mentioned above
\cite{ee}.

The untwisted sector can be obtained by keeping the fields invariant
under the projection.
It is not difficult to see that the ten-dimensional metric and dilaton
survive the projection, while the gauge boson is projected out.
Also the three-form is projected out, while the two-form survives.
Half of the fermions survive, a Majorana-Weyl gravitino and a
Mayorana-Weyl fermion of opposite chirality.
Thus, in the massless spectrum, we are left with the N=1 supergravity
multiplet.
We do know by now that this theory is anomalous in ten dimensions.
We must have some ``twisted sector" which should arrange itself to
cancel the anomalies.
As we discussed in the section on orbifolds, $S^1/Z_2$ is a line
segment, with the fixed-points $0,\pi$ at the boundary. The
fixed-planes are two copies of
ten-dimensional flat space.
States coming from the twisted sector must be localized on these
planes.
We have also a symmetry exchanging the fixed planes, so we expect
isomorphic massless content coming from the two fixed planes.
It can also be shown, that half of the anomalous variation is localized
at one fixed plane and the other half at the other.
The only N=1 multiplets which can cancel the anomaly symmetrically, are
vector multiplets, and we must have 248 of them at each fixed plane.
The possible anomaly free groups satisfying this constraint are $\rm
E_8\times E_8$
and U(1)$^{496}$.
Since there is no known string theory associated with the second
possibility,
it is natural to assume that we have obtained the E$_8\times$E$_8$
heterotic string theory.
A similar argument to that of the previous section shows that that
there is a relation similar to (\ref{604}) between the radius of the
orbifold and the heterotic coupling.
In the perturbative heterotic string, the two ten-dimensional planes
are on top of each other and they move further apart as the coupling
grows.

The M-theory membrane survives in the orbifold only if one of its
dimensions is wound around the $S^1/Z_2$. It provides the perturbative
heterotic string.
On the other hand, the five-brane survives, and cannot wind around the
orbifold direction. It provides the heterotic NS5-brane.
This is in accord with what we would expect from the heterotic string.
Upon compactification to four-dimensions, the NS5-brane will give rise
to magnetically charged point-like states (monopoles).

\renewcommand{\theequation}{\thesection.\arabic{equation}}
\section{Self-duality of the type-IIB string.}
\setcounter{equation}{0}

As described in section \ref{RRR}, the type-IIB theory in ten
dimensions
contains the following forms:

$\bullet$ The NS-NS two-form $B^1$. It couples electrically to the
perturbative
type-IIB string (which we will call for later convenience the (1,0)
string) and magnetically to a five-brane.

$\bullet$ The R-R scalar. It is a zero-form (there is a Peccei-Quinn
symmetry associated with it) and couples electrically to a (-1)-brane.
Strictly speaking this is an instanton whose ``world-volume" is a point
in spacetime.
It also couples magnetically to a seven-brane.

$\bullet$ The R-R two-form $B^2$. It couples electrically to a (0,1)
string
(distinct from the perturbative type-II string) and magnetically to
another (0,1) five-brane.

$\bullet$ The self-dual four-form. It couples to a self-dual
three-brane.

The theory is chiral but anomaly-free as we will see later on.
The self-duality
condition
implies that the field strength $F$ of the four-form is equal to its
dual.
This equation cannot be obtained from a covariant action.
Consequently, for type-IIB supergravity, the best we can do is to write
down the
equations of motion \cite{IIB}.

There is an SL(2,$\R$) global invariance in this theory which
transforms
the antisymmetric tensor and scalar doublets (the metric as well as
the four-form are invariant).
We will denote by $\phi$ the dilaton which comes from the
($NS-\overline{NS}$)
sector and by $\chi$ the scalar that comes from the ($R-\bar R$)
sector.
Define the complex scalar
\be
S=\chi+ie^{-\phi/2}
\,.\label{303}\ee
Then, SL(2,$\R)$ acts by fractional transformations on $S$ and linearly
on $B^i$
\be
S \to {aS+b\over cS+d}\;\;\;,\;\;\; \left(\matrix{B^N_{\mu\nu}\cr
B^{R}_{\mu\nu}\cr}\right)\to \left(\matrix{d&-c\cr -b&a\cr}\right)
 \left(\matrix{B^N_{\mu\nu}\cr
B^{R}_{\mu\nu}\cr}\right)
\,,\label{304}\ee
where $a,b,c,d$ are real with $ad-bc=1$.
$B^N$ is the NS-NS antisymmetric tensor while $B^R$ is the R-R
antisymmetric tensor.
When we set the four-form to zero, the rest of the equations of motion
can be obtained from the following action
\be
S^{IIB}={1\over 2\kappa^2}\int d^{10}x\sqrt{-\det g}\left[R-{1\over 2}
{\partial S\partial\bar S\over S_2^2}
-{1\over 12}{|H^R+ SH^N|^2\over S_2}\right]
\,,\label{IIB}\ee
where $H$ stands for the field strength of the antisymmetric tensors.
Obviously (\ref{IIB}) is SL(2,$\R$) invariant.

Its SL(2,$\Z$) subgroup was conjectured \cite{sch,ht} to be an exact
non-perturbative symmetry.

There is a (charge-one)  BPS instanton solution in type-IIB theory
given by the following configuration \cite{Din}
\be
e^{\phi/2}=\l+{c\over r^8}\;\;\;,\;\;\;\chi=\chi_0+i{c\over \l
(\l r^8+c)} \,,\label{607}\ee
where $r=|x-x_0|$, $x_0^{\m}$ being the position of the instanton, $\l$
is the string coupling far away from the instanton, $c=\pi\sqrt{\pi}$
is fixed by the requirement that the solution has minimal instanton
number and the other expectation values are trivial.

There is also a fundamental string solution which is charged under
$B^1$ (the (1,0) string), found in \cite{fstring}.
It has a singularity at the core, which is interpreted as a source for
the fundamental type-IIB string.
Acting with $S\to -1/S$ transformation on this solution we obtain
\cite{sch}
a solitonic string solution (the (0,1) string) that is charged under
the RR antisymmetric tensor $B^2$.
It is  given by the following configuration \cite{sch}
\be
ds^2=A(r)^{-3/4}[-(dx^0)^2+(dx^1)^2]+A(r)^{1/4}dy\cdot dy \;\;\;,\;\;\;
S=\chi_0+i{e^{-\phi_0/2}\over \sqrt{A(r)}}
\,,\label{608}\ee
\be
B^1=0\;\;\;,\;\;\;B^2_{01}={1\over \sqrt{\Delta} A(r)}
\,,\label{609}\ee
where
\be
A(r)=1+{Q\sqrt{\Delta}\over 3r^6}\;\;\;,\;\;\;Q={3\kappa^2 T\over
\pi^4}\;\;\;,\;\;\;
\Delta=e^{\phi_0/2}\left[\chi_0^2+e^{-\phi_0}\right]
\,.\label{610}\ee
$\kappa$ is Newton's constant and $T=1/(2\pi\a')$ is the tension of the
perturbative
type-IIB string.
The tension of the (0,1) string can be calculated to be
\be
\tilde T=T\sqrt{\Delta}
\,.\label{611}\ee
In the perturbative regime, $e^{\phi_0}\to 0$, $\tilde T\sim
Te^{-\phi_0/4}$ is large, and the (0,1) string is very stiff.
Its vibrating modes cannot be seen in perturbation theory.
However, at strong coupling, its fluctuations become the relevant low
energy
modes.
Acting further by SL(2,$\Z)$ transformations we can generate a
multiplet
of (p,q) strings with p,q relatively prime.
If such solitons are added to the perturbative theory, the continuous
SL(2,$\R$) symmetry is broken to SL(2,$\Z$).
All the (p,q) strings have a common massless spectrum given by the
type-IIB supergravity content. Their massive excitations are distinct.
Their string tension is given by
\be
T_{p,q}=T{|p+qS|^2\over S_2}\;\;.
\label{pq}\ee

By compactifying the type-IIB theory on a circle of radius $R_B$, it
becomes equivalent to the IIA theory compactified on a circle.
On the other hand, the nine-dimensional type-IIA theory is M-theory
compactified on a two-torus.

{}From the type IIB point of view, wrapping (p,q) strings around the
tenth dimension provides a spectrum of particles in nine-dimensions
with masses
\be
M^2_B={m^2\over R_B^2}+(2\pi R_BnT_{p,q})^2+4\pi T_{p,q}(N_L+N_R)\;\;,
\label{pq1}\ee
where m is the Kaluza-Klein  momentum integer, n the winding number
and $N_{L,R}$ the string oscillator numbers.
The matching condition is $N_L-N_R=mn$ and BPS states are obtained for
$N_{L}=0$ or $N_R=0$.
Thus, we obtain the following BPS spectrum
\be
\left. M^2_B\right|_{BPS}=\left({m\over R_B}+2\pi
R_BnT_{p,q}\right)^2\;\;.
\label{pq2}\ee
Since an arbitrary pair of integers $(n_1,n_2)$ can be written as
$n(p,q)$
where $n$ is the greatest common divisor and p,q are relatively prime
we can rewrite the BPS mass formula above as
\be
\left. M^2_B\right|_{BPS}=\left({m\over R_B}+2\pi
R_BT{|n_1+n_2S|^2\over S_2}\right)^2\;\;.
\label{pq3}\ee

In M-theory, compactified on a two-torus with area $A_{11}$ and modulus
$\tau$,
we have two types of (point-like) BPS states in nine dimensions:
KK states with mass $(2\pi)^2|n_1+n_2\tau|^2/(\tau_2A_{11})$ as well as
states
that are obtained by wrapping the M-theory membrane m times around the
two torus, with mass $(m A_{11} T_{11})^2 $, where $T_{11}$ is the
tension of the membrane.
We can also write $R_{11}$ that becomes the IIA coupling as
$R_{11}=A_{11}/(4\pi^2\tau_2)$.
Thus, the BPS spectrum is
\be
M_{11}^2=(m(2\pi R_{11})\tau_2 T_{11})^2+{|n_1+n_2\tau|^2\over
R_{11}^2\tau_2^2}+\cdots\;\;\;,
\label{pq4}\ee
where the dots are mixing terms that we cannot calculate.
The two BPS mass spectra should be related by $M_{B}=\b M_B$, where
$\b\not =1$
since the masses are measured in different units in the two theories.
Comparing, we obtain
\be
S=\tau\;\;\;,\;\;\;{1\over R_B^2}=T T_{11}A_{11}^{3/2}\;\;\;,\;\;\;
\beta=2\pi R_{11}{\sqrt{\tau_2}T_{11}\over T}\;\;.
\label{pq5}\ee
An outcome of this is the calculation of the M-theory membrane tension
$T_{11}$ in terms of string data.

\renewcommand{\theequation}{\thesection.\arabic{equation}}
\section{D-branes are the type-II RR charged states.}
\setcounter{equation}{0}

We have seen in section \ref{hetI} that D-branes defined by imposing
Dirichlet boundary conditions on some of the string coordinates
provided non-perturbative extended solitons required by heterotic-type
I string duality.

Similar D-branes can be also constructed in type-II string theory,
the only difference being that here, there is no orientifold
projection.
Also, open string fluctuations around them cannot have Neumann (free)
end-points.
As we will see, such D-branes will provide all RR charged states
required by the non-perturbative dualities of type-II string theory.

In the type-IIA theory we have seen that there are  (in principle)
allowed
 RR charged $p$-branes with  $p=0,2,4,6,8$, while in the type-IIB
$p=-1,1,3,5,7$.
D-branes can be constructed with a number of coordinates having
D-boundary conditions being $9-p=1,2,\ldots,10$, which precisely
matches the full allowed $p$-brane spectrum of type-II theories.
The important question is: are such D-branes charged under RR forms?

\begin{figure}
\begin{center}
\leavevmode
\epsfbox{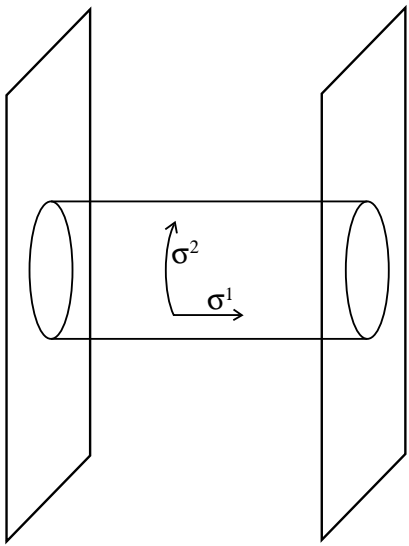}
\caption[]{D-branes interacting via the tree-level exchange of a closed
string.}
\label{f22}
\end{center}\end{figure}

To answer this question, we will have to study the tree-level
interaction
of two parallel D$p$-branes via the exchange of a closed string
\cite{Dp},
depicted schematically in Fig. \ref{f22}.
For this interpretation time runs horizontally.
However, if we take time to run vertically, then, the same diagram can
be interpreted as a (one-loop) vacuum fluctuation of open strings with
their
end-points attached to the D-branes.
In this second picture we can calculate this  diagram  to be
\be
{\cal A}=2V_{p+1}\int {d^{p+1}k\over
(2\pi)^{p+1}}\int_{0}^{\infty}{dt\over 2t}
e^{-2\pi\a' tk^2-t{|Y|^2\over 2\pi\a'}}{1\over \eta^{12}(it)}{1\over 2}
\sum_{a,b}(-1)^{a+b+ab}\th^4[^a_b](it)
\label{612}\ee
$$=2V_{p+1}\int_{0}^{\infty}{dt\over 2t}(8\pi^2\a't)^{-{p+1\over 2}}
e^{-t{|Y|^2\over 2\pi\a'}}{1\over \eta^{12}(it)}{1\over 2}
\sum_{a,b=0}^1(-1)^{a+b+ab}\th^4[^a_b](it)\,.$$

$V_{p+1}$ is the world-volume of the $p$-brane, the factor of two is
because
of the two end-points, $|Y|^2$ is the distance between the D-branes.
Of course the total result is zero, because of the $\th$-identity.
This reflects the fact that the D-branes are BPS states and exert no
static force on each other.
However, our purpose is to disentangle the contributions of the various
intermediate massless states in the closed string channel.
This can be obtained by taking the leading $t\to 0$ behavior of the
integrand.
In order to do this, we have to perform a modular transformation $t\to
1/t$
in the $\th$- and $\eta$-functions. We obtain
\be
\left.{\cal A}\right|^{\rm closed~~string}_{\rm massless}=8(1-1)V_{p+1}
\int_{0}^{\infty}{dt\over t}(8\pi^2\a' t)^{-{p+1\over
2}}~t^4~e^{-{t|Y|^2
\over 2\pi\a'}}
\label{613}\ee
$$=2\pi(1-1)V_{p+1}(4\pi^2\a')^{3-p}G_{9-p}(|Y|)
$$
where
\be
G_d(|Y|)={1\over 4\pi^{d/2}}\int_0^{\infty}{dt\over
t^{(4-d)/2}}e^{-t|Y|^2}
\label{614}\ee
is the massless scalar propagator in $d$ dimensions.
The $(1-1)$ comes from the NS-NS and R-R sectors respectively.
Now consider the RR forms coupled to p-branes with action
\be
S={\a_p\over 2}\int F_{p+2}\;^*F_{p+2}+iT_{p}\int_{\rm branes}A_{p+1}
\,,\label{615}\ee
with $F_{p+2}=dA_{p+1}$.
Using this action, the same amplitude for exchange of $A_{p+1}$ between
two D-branes at distance $|Y|$ in the transverse space of dimension
$10-(p+1)=9-p$ is given by
\be
\left.{\cal A}\right|_{\rm field~~theory}={(iT_{p})^2\over
\a_{p}}V_{p+1}G_{9-p}(|Y|)
\,,\label{616}\ee
where the factor of volume is there since the RR field can be absorbed
or emitted at any point in the world-volume of the D-brane.
Matching with the string calculation we obtain
\be
{T^2_p\over \a_p}=2\pi(4\pi^2\a')^{3-p}
\,.\label{617}\ee
We will  now look at the DNT quantization condition which, with our
normalization
of the RR forms, and $D=10$ becomes
\be
{T_pT_{6-p}\over \a_p}=2\pi n
\,.\label{618}\ee
{}From (\ref{617}) we can verify directly that D-branes satisfy this
quantization condition for the minimum quantum $n=1$!

Thus, we are led to accept that D-branes, with a nice (open) CFT
description of their fluctuations, describe non-perturbative extended
BPS states of the
type-II string carrying non-trivial RR charge.

We will now describe a uniform normalization of the D-brane
tensions.
Our starting point is the type-IIA ten-dimensional effective action
(\ref{596}).
The gravitational coupling $\kappa_{10}$ is given in terms of $\a'$ as
\be
2\kappa_{10}^{2}=(2\pi)^7\a'^4\;\;.
\label{nor1}\ee
We will also normalize all forms so that their kinetic terms
are $(1/4\kappa_{10}^2)\int d^{10}x F\otimes ^*F$.
This corresponds to $\a_{p}=1/(2\kappa_{10}^2)$.
We will define also the tensions of various p-branes via their
world-volume action of the form
\be
S_{p}=-T_p\int_{W_{p+1}}d^{p+1}\xi ~e^{-\Phi/2}\sqrt{{\rm det} \hat
G}-iT_p\int
{}~A_{p+1}\;\;\;,
\label{nor2}\ee
where $\hat G$ is the induced metric on the world-volume
\be
\hat G_{\a\b}=G_{\m\n}{\pd X^{\mu}\over \pd \xi^{\a}}{\pd X^{\n}\over
\pd \xi^{\b}}
\label{nor3}\ee
and
\be
\int A_{p+1}={1\over (p+1)!}\int d^{p+1}\xi
{}~A_{\m_{1}\cdots\m_{p+1}}{\pd X^{\m_1}\over \pd \xi^{\a_1}}\cdots
{\pd X^{\m_{p+1}}\over \pd \xi^{\a_{p+1}}}\e^{\a_1\cdots\a_{p+1}}\;\;.
\label{nor4}\ee
The dilaton dependence will be explained in the next section.
The DNT quantization condition in (\ref{618}) becomes
\be
2\kappa_{10}^2 T_{p}T_{6-p}=2\pi n\;\;,
\ee
while (\ref{617}) and (\ref{nor1}) give
\be
T_p={1\over (2\pi )^p (\a')^{(p+1)/2}}\;\;.
\label{nor5}\ee
We have obtained the IIA theory from the reduction of
eleven-dimensional supergravity on a circle of volume $2\pi R_{11}=2\pi
\sqrt{a'}e^{\Phi/3}$.
Consequently, the M-theory gravitational constant is
\be
2\kappa_{11}^2=(2\pi)^{8}(\a')^{9/2}\;\;.
\label{nor6}\ee
The M-theory membrane, upon compactification of M-theory on a circle,
becomes the type-IIA D2-brane.
Thus, its
tension $T^M_2$ should be equal to the D2-brane tension,
\be
T^M_2=T_2={1\over (2\pi)^2(\a')^{3/2}}\;\;.
\label{nor7}\ee
Consider now the M-theory five-brane. It has a tension $T^M_5$ that can
be computed from the DNT quantization condition
\be
2\kappa_{11}^2 T_2^MT^M_5=2\pi\;\;\;\to\;\;\;\;T^M_5={1\over
(2\pi)^5(\a')^3}\;.
\label{nor8}\ee
On the other hand, wrapping one of the coordinates of the M5-brane
around the circle should produce the D4-brane and we can confirm that
\be
2\pi\sqrt{\a'}T^M_5=T_4\;.
\label{nor9}\ee

\renewcommand{\theequation}{\thesection.\arabic{equation}}
\section{D-brane actions}
\setcounter{equation}{0}

We will now derive the massless fluctuations of a single Dp-brane.
This parallels our detailed discussion of the type-I D1-brane.
The  difference here is that the open string fluctuations cannot have
free
ends\footnote{Free end-points are interpreted as 9-branes and there are
none
in type-II string theory.}.
Thus, only the DD sector is relevant.
Also there is no orientifold projection.
In the NS sector, the massless bosonic states are a (p+1)-vector,
$A_{\m}$
corresponding to the state $b^{\mu}_{-1/2}|p\rangle$ and 9-p scalars,
$X^I$ corresponding to the states $b^I_{-1/2}|p\rangle$.
The $X^I$ represent the position coordinates of the Dp-brane in
transverse space.
These are the states we would obtain by reducing a ten-dimensional
vector
to p+1 dimensions.
Similarly, from the R sector we obtain world-volume fermions which are
the reduction of a ten-dimensional gaugino to (p+1) dimensions.
In total we obtain the reduction of a ten-dimensional U(1) vector
multiplet
to p+1 dimensions.
The world-volume supersymmetry has 16 conserved supercharges. Thus, the
Dp-brane broke half-of the original supersymmetry as expected.

In order to calculate the world-volume action, we would have to
calculate scattering of the massless states of the world-volume theory.
The leading contribution comes from the disk diagram and is thus
weighted
with a factor of $e^{-\Phi/2}$.
The calculation is similar with the calculation of the effective action
in the ten-dimensional open oriented string theory.
The result there is the Born-Infeld action for the gauge field
\cite{bi}
\be
S_{BI}=\int d^{10}x~ e^{-\Phi/2}\sqrt{\det(\delta_{\m\n}+2\pi \a'
F_{\m\n})}\;\;.
\label{nor10}\ee
Dimensionally reducing this action, we obtain the relevant Dp-brane
action
from the disk.
There is a coupling to the spacetime background metric which gives the
induced metric, (\ref{nor3}).
There is also a coupling to the spacetime NS antisymmetric tensor. This
can be seen as follows.
The closed string coupling to $B_{\m\n}$ and the vector $A_{\mu}$ can
be summarized in
\be
S_{B}={i\over 2\pi\a'}\int_{M_2}
d^2\xi ~\e^{\a\b}B_{\m\n}\pd_{a}x^{\mu}\pd_{\b}x^{\nu}-{i\over
2}\int_{B_1}ds ~A_{\mu}\pd_s x^{\m}\;\;,
\label{nor11}\ee
where $M_2$ is the two-dimensional world-sheet with one-dimensional
boundary
$B_1$.
Under a gauge transformation $\delta
B_{\m\n}=\pd_{\mu}\Lambda_{\n}-\pd_{\n}\Lambda_{\m}$, the action above
changes
by a boundary term,
\be
\delta S_{B}={i\over \pi\a'}\int_{B_1} ds ~\Lambda_{\m}\pd_s
x^{\m}\;\;.
\label{nor12}\ee
To reinstate gauge invariance, the vector $A_{\m}$ has to transform as
$\delta A_{\mu}={1\over 2\pi\a'}\Lambda_{\mu}$.
Thus, the gauge invariant combination is
\be
{\cal F}_{\mu\nu}=2\pi \a' F_{\m\n}-B_{\mu\nu}=2\pi
\a'(\pd_{\mu}A_{\n}-\pd_{\n}A_{\m})-B_{\mu\nu}\;\;.
\label{nor13}\ee
We can now sumarize the leading order Dp-brane action as
\be
S_{p}=-T_p\int_{W_{p+1}}d^{p+1}\xi ~e^{-\Phi/2}\sqrt{{\rm det}( \hat
G+{\cal F})}-iT_p\int ~A_{p+1}\;\;.
\label{nor14}\ee
As we have seen in the previous section, the CP-odd term in the action
comes from the next diagram, the annulus.
There are however more CP-odd couplings coming from the annulus that
involve q-forms with q$<$p.
Their appearance is due to cancellation of anomalies, and we refer the
reader to \cite{inflow} for a detailed discussion.
We will present here the result.
It involves the roof-genus  $\hat I_{1/2}(R)$ and the Chern character.
Thus,
(\ref{nor14}) is extended to
\be
S_{p}=-T_p\int_{W_{p+1}}d^{p+1}\xi ~e^{-\Phi/2}\sqrt{{\rm det}( \hat
G+{\cal F})}-iT_p\int ~A\wedge Tr[e^{i{\cal F}/2\pi}]\sqrt{\hat
I_{1/2}(R)}\;\;,
\label{nor15}\ee
where $A$ stands for a formal sum of all RR forms, and the integration
picks up the (p+1)-form in the sum.

As an example we will consider the action of the  D1-string of type-IIB
theory.
The relevant forms that couple here is the RR two-form $B^R_{\m\n}$
as well as the RR scalar (zero-form) $S_1$.
The action is
\be
S_{1}=-{1\over 2\pi \a'}\left[\int d^{2}\xi ~{|S|\over
\sqrt{S_2}}~\sqrt{{\rm det}( \hat G+{\cal F})}+i\int (B^N+{iS_1\over
2\pi}{\cal F})\right]\;\;,
\label{nor16}\ee
where $e^{-\Phi/2}=S_2$.
Note that ${|S|\over \sqrt{S_2}}=e^{-\Phi/2}$ when $S_1=0$.

We will now consider the effect of T-duality transformations on the
Dp-branes.
Consider the type-II theory with $x^9$ compactified on a circle of
radius R.
As we have mentioned earlier, the effect of a T-duality transformation
on open strings is to interchange N and D boundary conditions.
Consider first a Dp-brane not wrapping around the circle.
This implies that one of its transverse coordinates (Dirichlet) is in
the compact direction.
Doing a T-duality transformation $R\to \a'/R$, would change the
boundary conditions along $X^9$ to Neumann and would produce a
D(p+1)-brane wrapping
around the circle of radius $\a'/R$.
Thus, the Dp-brane has been transformed into a D(p+1)-brane.
The original Dp-brane action contains $T_p\int d^{p+1}\xi~e^{-\Phi/2}$.
The dilaton transforms under duality as
\be
e^{-\Phi/2}\to {\sqrt{\a'}\over R}e^{-\Phi/2}\;\;.
\ee
Consequently, $T_p\sqrt{\a'}/R=T_{p+1}(2\pi \a'/R)$ and we obtain
\be
T_{p+1}={T_p\over 2\pi \sqrt{\a'}}\;\;,
\label{nor17}\ee
which is in agreement with (\ref{nor5}).

On the other hand, if the Dp-brane was wrapped around the compact
direction, T-duality transforms it into a D(p-1)-brane.
This action of T-duality on the various D-branes is a powerful tool
for investigating non-perturbative physics due to them.

So far, we have discussed a single Dp-brane, interacting with the
background type-II fields. An obvious question is: what happens when we
have more
than one parallel Dp-branes?
Consider first the case where we have N Dp-branes being at the same
point in transverse space. Then, the only difference
in the previous analysis, is to include a Chan-Patton factor
$i=1,2,\cdots ,N$
at the open string end-points.
We now have $N^2$ massless vector states,
$b^{\mu}_{-1/2}|p;i,j\rangle$.
Going through the same procedure as before, we will find that the
massless fluctuations are described by the dimensional reduction of the
ten-dimensional
N=1 U(N) Yang-Mills multiplet on the world-volume of the brane (we have
oriented open strings here).
The U(1) factor of U(N) describes the overall center of mass of the
system.
If we take one of the Dp-branes and we separate it from the rest, the
open strings stretching between it and the rest N-1 of the branes,
acquire a mass-gap
(non-trivial tension), and the massless vectors have a gauge group
which is
$\rm U(N-1)\times U(1)$.
In terms of the world-sheet theory, this is an ordinary Higgs effect.
For generic positions of the Dp-branes, the gauge group is $\rm
U(1)^N$.
The scalars that described the individual positions become now U(N)
matrices.
The world-volume action has a non-abelian generalization. In
particular,
to lowest order, it is the dimensional reduction of U(N)
ten-dimensional Yang-Mills
\be
S^N_{p}=-T_p Str\int_{W_{p+1}}d^{p+1}\xi ~e^{-\Phi/2}
(F_{\m\n}^2+2F_{\m I}^2+F_{IJ}^2)\;\;,
\label{nor18}\ee
where
\be
F_{\m\n}=\pd_{\m}A_{\n}-\pd_{\n}A_{\m}+[A_{\mu},A_{\n}]\;\;,
\label{nor19}\ee
\be
F_{\mu I}=\pd_{\m}X^I+[A_{\m},X^I]\;\;\;,\;\;\;F_{IJ}=[X^I,X^J]\;.
\label{nor20}\ee
Both $A_{\mu}$ and $X^I$ are U(N) matrices.
At the minimum of the potential, the matrices $X^I$ are commuting, and
can be simultaneously diagonalized. Their eigenvalues can be
interpreted
as the coordinates of the N Dp-branes.

One very interesting application of D-branes is the following.
D-branes wrapped around compact manifolds produce point-like RR charged
particles in lower dimensions.
Such particles have an effective description as microscopic black
holes.
Using D-brane techniques, their multiplicity can be computed for fixed
charge and mass. It can be shown that this multiplicity agrees to
leading order
with the Bekenstein-Hawking entropy formula for classical black holes
\cite{vs}. The interested reader may consult \cite{mal} for a review.

\renewcommand{\theequation}{\thesection.\arabic{equation}}
\section{Heterotic/Type-II duality in six and four Dimensions}
\setcounter{equation}{0}

There is another non-trivial duality relation that we are going to
discuss
in some detail: that of the heterotic string compactified to six
dimensions on $T^4$ and the type-IIA string compactified on K3.
Both theories have N=2 supersymmetry in six dimensions.
Both theories have the same massless spectrum, containing the N=2
supergravity multiplet and twenty vector multiplets.

The six-dimensional tree-level heterotic effective action in the
$\s$-model frame is
\be
S^{\rm heterotic}_{D}=\int ~d^Dx~\sqrt{-{\rm
det}~G}e^{-\Phi}\left[R+\pd^{\m}\Phi\pd_{\m}\Phi
-{1\over 12} \hat H^{\m\n\rho}\hat H_{\m\n\rho}-\right.
\label{348}\ee
$$\left.-{1\over 4}(\hat M^{-1})_{ij}
F^{i}_{\m\n}F^{j\m\n}+{1\over 8}Tr(\pd_{\m} \hat M\pd^{\m}
\hat M^{-1})\right]\,,
$$
where $i=1,2,\dots,36-2D$ and
\be
\hat H_{\m\n\rho}=\pd_{\m}B_{\n\rho}-{1\over 2}L_{ij}
A^i_{\m}F^j_{\n\rho}
+{\rm cyclic}
\,.\label{350}\ee
The moduli scalar matrix $\hat M$ is,
\be
M=\left(\matrix{G^{-1}& G^{-1}C &G^{-1}Y^{t}\cr
C^{t}G^{-1}&G+C^{t}G^{-1}C+Y^{t}Y&C^{t}G^{-1}Y^{t}+Y^{t}\cr
YG^{-1}&YG^{-1}C+Y&{\bf 1}_{16}+YG^{-1}Y^{t}\cr}\right)
\,,\label{B5}\ee
where ${\bf 1}_{16}$ is the sixteen-dimensional unit matrix and
\be
C_{\a\b}=B_{\a\b}-{1\over 2}Y^{I}_{\a}Y^{I}_{\b}
\,.\label{B6}\ee

Going to the Einstein frame by $G_{\m\n}\to e^{\Phi/2}G_{\m\n}$, we
obtain
\be
S^{\rm het}_{D=6}=\int ~d^6x~\sqrt{-G}\left[R-{1\over
4}\pd^{\m}\Phi\pd_{\m}\Phi
-{e^{-\Phi}\over 12}\hat H^{\m\n\rho}\hat H_{\m\n\rho}-\right.
\label{6188}\ee
$$\left.-{e^{-{\Phi\over 2}}\over 4}(\hat M^{-1})_{ij}
F^{i}_{\m\n}F^{j\m\n}+{1\over 8}Tr(\pd_{\m} \hat M\pd^{\m}
\hat M^{-1})\right]\,.
$$

The tree-level type-IIA effective action in the $\s$-model frame is
\be
S^{IIA}_{K3}=\int
d^6x\sqrt{-{\rm det}~G_6}e^{-\Phi}\left[R+\na^{\m}\Phi\na_{\m}\Phi
-{1\over 12}H^{\m\n\rho}H_{\m\n\rho}+\right.
\label{377}\ee
$$\left.+{1\over 8}Tr(\pd_{\m} \hat M\pd^{\m}\hat
M^{-1})\right]-{1\over
4}\int
d^6x\sqrt{-{\rm det}~G}(\hat M^{-1})_{IJ}
F^{I}_{\m\n}F^{J\m\n}+{1\over 16}\int d^6x
\e^{\m\n\rho\s\tau\upsilon}B_{\m\n}F^{I}_{\rho\s}\hat L_{IJ}
F^{J}_{\tau\upsilon}\,,
$$
where $I=1,2,\dots,24$.

Going again to the Einstein frame we obtain

\be
S^{IIA}_{D=6}=\int d^6x\sqrt{-G}\left[R-{1\over
4}\pd^{\m}\Phi\pd_{\m}\Phi
-{1\over 12}e^{-\Phi}H^{\m\n\rho}H_{\m\n\rho}-\right.
\label{619}\ee
$$\left.-{1\over 4}e^{\Phi/2}(\hat M^{-1})_{ij}
F^{i}_{\m\n}F^{j\m\n}+{1\over
8}Tr(\pd_{\m}\hat M\pd^{\m}\hat M^{-1})\right]+{1\over 16}\int d^6x
\e^{\m\n\rho\s\tau\varepsilon}B_{\m\n}F^{i}_{\rho\s}\hat
L_{ij}F^{j}_{\tau\varepsilon}\;,
$$
where $\hat L$ is the O(4,20) invariant metric.
Notice the following differences: The heterotic $\hat H_{\m\n\rho}$
contains the Chern-Simons term (\ref{350}) while the type-IIA one
doesn't.
The type-IIA action instead contains a parity-odd term coupling the
gauge
fields and $B_{\m\n}$.
Both effective actions have a continuous O(4,20,$\R$) symmetry which is
broken in the string theory to the T-duality group O(4,20,$\Z$).

We will denote by a prime the fields of the type-IIA theory (Einstein
frame)
and without a prime those of the heterotic theory.

\vskip .4cm
\noindent\hrulefill
\vskip .4cm

{\large\bf Exercise}. Derive the equations of motion stemming from the
actions (\ref{6188}) and (\ref{619}). Show that the two sets of
equations of motion  are equivalent via the following (duality)
transformations

\be
\Phi'=-\Phi\;\;\;,\;\;\;G'_{\m\n}=G_{\m\n}\;\;\;,\;\;\;\hat M'=\hat
M\;\;\;,\;\;\;
A'^i_{\mu}=A^{i}_{\mu}
\,,\label{620}\ee
\be
e^{-\Phi}\hat H_{\m\n\rho}={1\over
6}{{\e_{\m\n\rho}}^{\s\tau\varepsilon}
\over\sqrt{-G}}H'_{\s\tau\varepsilon}
\,,\label{621}\ee
where the data on the right-hand side are evaluated in the type-IIA
theory.

\vskip .4cm
\noindent\hrulefill
\vskip .4cm

There is a way to see some indication of this duality by considering
the compactification of M-theory on $S^1\times$K3 which is equivalent
to type-IIA on K3.
As we have seen in a previous section, all vectors descend from the RR
one-
and three-forms of the ten-dimensional type-IIA theory, and these
descend
from the three-form of M-theory to which the membrane and five-brane
couple.
The membrane wrapped around $S^1$ would give a string in six
dimensions.
Like in ten dimensions, this is the perturbative type-IIA string.
There is another string however, obtained by wrapping the five-brane
around the whole K3. This is the heterotic string \cite{ks}.

There is further evidence for this duality.
The effective action of type-IIA theory on K3 has a string solution
singular at the core.
The zero mode structure of the string is similar to the perturbative
type-IIA string.
There is also a string solution which is regular at the core. This is a
solitonic string and analysis of its zero modes indicates that it has
the same (chiral) word-sheet structure as the heterotic
string\footnote{We have seen a similar phenomenon already in the case
of the D1-string of type I string theory.}.
The string-string duality map (\ref{620}-\ref{621}) exchanges the roles
of the two strings.
The type-IIA string now becomes regular (solitonic), while the
heterotic
string solution becomes singular.

We will now compactify further both theories on a two-torus down to
four dimensions and examine the consequences of the duality.
In both cases we use the standard Kaluza-Klein ansatz described in
Appendix A.
The four-dimensional dilaton becomes as usual
\be
\phi=\Phi-{1\over 2}\log[\det G_{\a\b}]
\,,\label{622}\ee
where $G_{\a\b}$ is the metric of $T^2$ and $B_{\a\b}=\e_{\a\b} B$ is
the antisymmetric tensor.
We obtain
\be
S_{D=4}^{het}=\int d^4 x
\sqrt{-g}e^{-\phi}\left[R+L_{B}+L_{\rm gauge}+L_{\rm
scalar}\right]
\,,\label{623}\ee
where
\be
L_{g+\phi}=R+\pd^{\m}\phi\pd_{\m}\phi
\,,\label{624}\ee
\be
L_{B}=-{1\over 12}H^{\m\n\rho}H_{\m\n\rho}
\,,\label{625}\ee
with
\be
H_{\m\n\rho}=\pd_{\m}B_{\n\rho}-{1\over
2}\left[B_{\m\a}F^{A,\a}_{\n\rho}+A^{\a}_{\m}
F^{B}_{a,\n\rho}+\hat L_{ij}A^{i}_{\mu}F^{j}_{\n\rho}\right]+{\rm
cyclic}
\label{626}
\ee
$$\equiv \pd_{\m}B_{\n\rho}-{1\over 2}
L_{IJ}A^{I}_{\m}F^{J}_{\n\rho}+{\rm cyclic}
\,.$$
The matrix
\be
L=\left(\matrix{ 0&0&1&0& {\vec 0}\cr
0&0&0&1&\vec 0\cr
1&0&0&0&\vec 0\cr
0&1&0&0&\vec 0\cr
\vec 0&\vec 0&\vec 0&\vec 0&\hat L\cr}\right)
\label{627}\ee
is the O(6,22) invariant metric.
Also
\be
C_{\a\b}=\e_{\a\b}B-{1\over 2}\hat L_{ij}Y^i_{\a}Y^j_{\b}
\,,\label{6666}\ee
so that
$$
L_{\rm gauge}=-{1\over
4}\left\{\left[(\hat M^{-1})_{ij}+\hat L_{ki}\hat
L_{lj}Y^{k}_{\a}G^{\a\b}
Y^{l}_{\b}\right]
\;\;F^{i}_{\m\n}F^{j,\m\n}+G^{\a\b}\;\;F^{B}_{\a,\m\n}
F_{B,\b}^{\m\n}+\right.
$$
\be
+\left[G_{\a\b}+C_{\g\a}G^{\g\d}C_{\d\b}+Y^{i}_{\a}(\hat
M^{-1})_{ij}
Y^{j}_{\b}\right]
\;\;F^{A,a}_{\m\n}F_{A}^{\b,\m\n}-2G^{\a\g}C_{\g\b}\;\;
F^{B}_{\a,\m\n}F^{A,\b,\m\n}-
\label{628}\ee
$$
\left.-2\hat L_{ij}Y^{i}_{\a}G^{\a\b}\;\;F^{j}_{\m\n}F^{B,\m\n}_{\b}
+2(Y^{i}_{\a}(\hat M^{-1})_{ij}+C_{\g\a}G^{\g\b}\hat
L_{ij}Y^{i}_{\b})\;\;
F^{a,A}_{\m\n}
F^{j,\m\n}\right\}
$$
$$
\equiv -{1\over 4}(M^{-1})_{IJ}F^{I}_{\m\n}F^{J,\m\n}\;\;,
$$
where the index I takes 28 values. For the scalars
$$
L_{\rm scalar}=\pd_{\m}\phi\pd^{\m}\phi+{1\over 8}Tr[\pd_{\m}\hat
M\pd^{\m}\hat M^{-1}]-{1\over
2}G^{\a\b}(\hat M^{-1})_{ij}
\pd_{\m}Y^{i}_{\a}\pd^{\m}Y^{j}_{\b}+
$$
\be
+{1\over 4}\pd_{\m}G_{\a\b}\pd^{\m}G^{\a\b}-{1\over 2det G}
\left[\pd_{\m}B+\e^{\a\b}\hat
L_{ij}Y^{i}_{\a}\pd_{\mu}Y^{j}_{\b}\right]
\left[\pd^{\m}B+\e^{\a\b}\hat
L_{ij}Y^{i}_{\a}\pd^{\mu}Y^{j}_{\b}\right]\label{629}
\ee
$$
=\pd_{\m}\phi\pd^{\m}\phi+{1\over 8}Tr[\pd_{\m}M\pd^{\m} M^{-1}]\,.
$$

We will now go to the standard axion basis in terms of the usual
duality transformation in four dimensions.
First we will go to the Einstein frame by
\be
g_{\m\n}\to e^{-\phi}g_{\m\n}
\,,\label{630}\ee
so that the action becomes
\be
S^{het,E}_{D=4}=\int
d^4x\sqrt{-g}\;\left[R-{1\over
2}\pd^{\m}\phi\pd_{\m}\phi
-{1\over 12}e^{-2\phi}H^{\m\n\rho}H_{\m\n\rho}-\right.
\label{631}\ee
$$\left.-{1\over 4}e^{-\phi}(M^{-1})_{IJ}
F^{I}_{\m\n}F^{J,\m\n}+{1\over 8}Tr(\pd_{\m}M\pd^{\m}
M^{-1})\right]
\,.$$

The axion is introduced as usual,
\be
e^{-2\phi}H_{\m\n\rho}={{\e_{\m\n\rho}}^{\s}\over \sqrt{-g}}\pd_{\s}a
\,.\label{632}\ee
The transformed equations come from the
following
action:
\be
\tilde S^{het}_{D=4}=\int
d^4x\sqrt{-g}\;\left[R-{1\over
2}\pd^{\m}\phi\pd_{\m}\phi
-{1\over 2}e^{2\phi}\pd^{\m}a\pd_{\m}a-{1\over 4}e^{-\phi}(
M^{-1})_{IJ}
F^{I}_{\m\n}F^{J,\m\n}\right.
\label{633}\ee
$$\left.+{1\over 4}a\;L_{IJ}F^{I}_{\m\n}\tilde
F^{J,\m\n}+{1\over 8}Tr(\pd_{\m}M\pd^{\m}M^{-1})\right]\;\;,
$$
where
\be
\tilde F^{\m\n}={1\over 2}{\e^{\m\n\rho\s}\over \sqrt{-g}}
F_{\rho\s}
\,.\label{634}\ee

Finally, defining the complex S field
\be
S=a+i\; e^{-\phi}
\,,\label{635}\ee
we obtain
\be
\tilde S^{het}_{D=4}=\int
d^4x\sqrt{-g}\;\left[R-{1\over 2}{\pd^{\m}S\pd_{\m}\bar
S\over
{\rm Im}S^2}-{1\over 4}{\rm Im}S(M^{-1})_{IJ}
F^{I}_{\m\n}F^{J,\m\n}\right.
\label{637}\ee
$$\left.+{1\over 4}{\rm Re}S\;L_{IJ}F^{I}_{\m\n}\tilde
F^{J,\m\n}+{1\over 8}Tr(\pd_{\m}M\pd^{\m}M^{-1})\right]
\,.$$

Now consider the type-IIA action (\ref{377}).
Going through the same procedure  and introducing the axion via
\be
e^{-2\phi}H_{\m\n\rho}={{\e_{\m\n\rho}}^{\s}\over \sqrt{-g}}
\left[\pd_{\s}a+{1\over
2}\hat L_{ij}Y^{i}_{\a}\d_{\s}Y^{j}_{\b}\e^{\a\b}\right]
\,,\label{641}\ee
we obtain the following four-dimensional  action in the Einstein frame
\be
\tilde S_{D=4}^{IIA}=\int d^4 x
\sqrt{-g}\left[R+L^{\rm even}_{\rm gauge}+L^{\rm odd}_{\rm
gauge}+L_{\rm scalar}\right]
\,,\label{638}\ee
with
\be
L^{\rm even}_{\rm gauge}=-{1\over 4}\int d^4x\sqrt{-g}\left[
e^{-\phi}G^{\a\b}
\left(F^{B}_{\a,\m\n}-B_{\a\g}F^{A,\g}_{\m\n}\right)
\left(F^{B,\m\n}_{\b}-B_{\a\d}F^{\d,\m\n}_{A}\right)+\right.
\label{639}\ee
$$\left.
+e^{-\phi}G_{\a\b}F^{A,\a}_{\m\n}F^{\b,\m\n}_{A}+\sqrt{{\rm
det}G_{\a\b}}(\hat M^{-1})_{ij}\left(F^{i}_{\m\n}+Y^{i}_{\a}
F^{A,\a}_{\m\n}\right)
\left(F^{j,\m\n}+Y^{j}_{\b}F^{\b,\m\n}_{A}\right)\right]\,,
$$

\be
L^{\rm odd}_{\rm gauge}={1\over 2}\int d^4
x\e^{\m\n\rho\s}\left[{1\over
4}a
F^{B}_{\a,\m\n}F^{A,\a}_{\rho\s}+{1\over
2}\e^{\a\b}\hat
L_{ij}Y^{i}_{\b}F^{B}_{\a,\m\n}\left(F^{j}_{\rho\s}+{1\over
2}Y^{j}_{\g}F^{A,\g}_{\rho\s}\right)
\right.
\label{640}\ee
$$\left.
-{1\over 8} \e^{\a\b}\hat L_{ij}B_{\a\b}
\left(F^{i}_{\m\n}+Y^{i}_{\g}F^{A,\g}_{\m\n}\right)\left(
F^{j}_{\rho\s}+Y^{j}_{\d}F^{A,\d}_{\rho\s}\right)\right]
$$\,,

\be
L_{\rm scalar}=-{1\over 2}(\pd\phi)^2+{1\over
4}\pd^{\m}G_{\a\b}\pd_{\m}
G^{\a\b}
-{1\over 2det G}\pd_{\m} B\pd^{\m} B+{1\over
8}Tr[\pd_{\m}\hat M\pd^{\m} \hat M^{-1}]+\label{642}\ee
$$
-{1\over 2}e^{2\phi}(\pd_{\m}a+{1\over 2}\hat
L_{ij}\e^{\a\b}Y^{i}_{\a}\pd^{\m}
Y^{j}_{\b})^2
-{1\over 2}e^{\phi}\sqrt{{\rm det}G_{\a\b}}(\hat M^{-1})_{ij}G^{\a\b}
\pd_{\m} Y^{i}_{\a}\pd^{\m} Y^{j}_{\b}
\,.$$

Now we will use unprimed fields to refer to the heterotic side and
primed ones for the type-II side.
We will now work out the implications of the six-dimensional
duality relations
(\ref{620},\ref{621}) in four dimensions.
{}From (\ref{620}), we obtain
\be
e^{-\phi}=\sqrt{{\rm det}G'_{\a\b}}\;\;\;,\;\;\;
e^{-\phi'}=\sqrt{{\rm det}G_{\a\b}}
\,,\label{643}\ee
\be
{G_{\a\b}\over \sqrt{{\rm det}G_{\a\b}}}={G'_{\a\b}\over \sqrt{{\rm
det}G'_{\a\b}}}\;\;\;,\;\;\;A'^{\a}_{\m}=A^{\a}_{\m}
\,,\label{644}\ee
\be
g_{\m\n}=g'_{\m\n}\;\;\;\;\;{\rm Einstein~~frame}
\,,\label{645}\ee
\be
\hat M'=\hat M\;\;\;,\;\;\;A^{i}_{\m}=A'^{i}_{\m}\;\;\;,\;\;\;
Y^{i}_{\a}=Y'^{i}_{\a}
\,.\label{646}\ee

Finally, the relation (\ref{621}) implies
\be
A=B'\;\;\;,\;\;\;A'=B\label{647}
\ee
and
\be
{1\over 2}{{\e_{\m\n}}^{\rho\s}\over
\sqrt{-g}}\e^{\a\b}F^{B'}_{\b,\rho\s}=
e^{-\phi}G^{\a\b}\left[F^{B}_{\b,\m\n}-C_{\b\g}F^{A,\g}_{\m\n}-\hat
L_{ij}
Y^{i}_{\b}F^{j}_{\m\n}\right]-{1\over 2}a{{\e_{\m\n}}^{\rho\s}\over
\sqrt{-g}}F^{A,\a}_{\rho\s}\,,\label{648}\ee
which is an electric-magnetic duality transformation on the $B_{\a,\m}$
gauge fields (see Appendix D).
It is easy to check that this duality maps the scalar heterotic terms
to the type-IIA ones and vice versa.

In the following, we will keep the 4 moduli of the two torus and the 16
Wilson lines $Y^i_{\a}$
In the heterotic case we will define the $T,U$ moduli of the torus
and the complex Wilson lines as
\be
W^i=W^i_1+iW_2^i=-Y^i_2+UY^i_1
\,,\label{649}\ee
\be
G_{\a\b}={T_2-{\sum_i(W^i_2)^2\over 2 U_2}\over
U_2}\left(\matrix{1&U_1\cr
U_1&|U|^2\cr}\right)\;\;\;,\;\;\;B=T_1-{\sum_i W_1^iW_2^i\over 2U_2}
\,.\label{650}\ee
Altogether we have the complex field S$\in$SU(1,1)/U(1) (\ref{635})
and the $T,U,W^i$ moduli $\in$ ${{\rm O(2,18)}\over {\rm O(2)\times
O(18)}}$.
Then the relevant scalar kinetic terms can be written as
\be
L^{het}_{\rm scalar}=-{1\over 2}\pd_{z^i}\pd_{\bar z^j}K(z_k,\bar z_k)~
\pd_{\m}z^i\pd^{\m}\bar z^j
\,,\label{651}\ee
where the K\"ahler potential is
\be
K=\log\left[S_2\left(T_2U_2-{1\over 2}\sum_i(W_2^i)^2\right)\right]
  \,.\label{652}\ee

In the type-IIA case the complex structure is different:
(\ref{649}) remains the same but
\be
G_{\a\b}={T_2\over U_2}\left(\matrix{1&U_1\cr
U_1&|U|^2\cr}\right)\;\;\;,\;\;\;B=T_1
\,.\label{653}\ee
Also
\be
S=a-{\sum_i W_1^iW_2^i\over 2U_2}+i(e^{-\phi}-{\sum_i(W^i_2)^2\over 2
U_2})
\,.\label{654}\ee
Here $T\in $SU(1,1)/U(1) and $S,U,W^i\in {{\rm O(2,18)\over O(2)\times
O(18)}}$.
In this language the duality transformations become
\be
S'=T\;\;\;,\;\;\;T'=S\;\;\;,\;\;\;U=U'\;\;\;,\;\;\;W^i=W'^i
\,.\label{655}\ee

In the type-IIA string, there is a SL(2,$\Z)$ $T$-duality symmetry
acting on T
by fractional transformations.
This is a good symmetry in perturbation theory.
We also expect it to be a good symmetry non-perturbatively, since  it
is a discrete remnant of a gauge symmetry and is not expected to be
broken by non-perturbative effects.
Then heterotic/type-II duality implies that there is an SL(2,$\Z)$
S-symmetry that acts on the coupling constant and the axion.
This is a non-perturbative symmetry from the point of view of the
heterotic string.
It acts as an electric magnetic duality on all the 28 gauge fields.
In the field theory limit it implies an S-duality symmetry for N=4
super Yang-Mills theory in four dimensions.

We will finally see how heterotic/type-II duality acts on the 28
electric and
28 magnetic charges.
Label the electric charges by a vector ($m_1,m_2,n_1,n_2,q^i$)
where $m_i$ are the momenta of the two torus, $n_i$ are the respective
winding numbers, and $q^i$ are the rest of the 24 charges.
For the magnetic charges we write the vector  ($\tilde m_1,\tilde
m_2,\tilde n_1,\tilde n_2,\tilde q^i$).
Because of (\ref{648}) we have the following duality map.
\be
\left(\matrix{m_1\cr m_2\cr n_1 \cr
n_2\cr q^i\cr}\right)
\to\left(\matrix{m_1\cr m_2\cr \tilde n_2 \cr-\tilde
n_1\cr q^i\cr}\right)\;\;,\;\;
\left(\matrix{\tilde m_1\cr \tilde m_2\cr \tilde n_1 \cr \tilde
n_2\cr\tilde q^i\cr}\right)
\to\left(\matrix{\tilde m_1 \cr \tilde m_2\cr -n_2\cr n_1\cr\tilde
q^i\cr}\right)
\,.\label{656}\ee
One can compute the spectrum of BPS multiplets both short and
intermediate.
The results of section 12 are useful in this respect.
\vskip .4cm
\noindent\hrulefill
\vskip .4cm

{\large\bf Exercise}. Find the BPS multiplicities on the heterotic
and type-IIA side in four dimensions.

\vskip .4cm
\noindent\hrulefill
\vskip .4cm

There are indirect quantitative tests of this duality.
Compactifying the heterotic string to four dimensions with N=2
supersymmetry
can be dual to the type-IIA string compactified on a CY manifold
of a special kind (K3 fibration over $P^1$) \cite{kv,fhsv,re}.
In the heterotic theory, the dilaton is in a vector multiplet.
Consequently,  the vector multiplet moduli space has perturbative and
non-perturbative corrections
while the hypermultiplet moduli space is exact.
In the dual type-II theory, the dilaton is in a hypermultiplet.
Consequently, the vector moduli space geometry has no corrections and
can be computed at tree-level.
Doing the duality map that should reproduce all quantum corrections to
the heterotic side.
This has been done in some examples, and in this way the one-loop
heterotic correction was obtained which agreed with the heterotic
computation.
Moreover, all instanton effects were obtained this way.
Taking the field theory limit and decoupling gravity, the
Seiberg-Witten solution was verified for N=2 gauge theory.
This procedure gives also a geometric interpretation of the
Seiberg-Witten solution.
A review of these developements can be found in \cite{lere}.

\section{Helicity string partition functions and multiplicities of BPS
states}
\renewcommand{\theequation}{\thesection.\arabic{equation}}
\setcounter{equation}{0}

We have seen in section 3 that BPS states are important ingredients in
non-perturbative dualities. The reason is that their special
properties,
most of the time, guarantee that such states survive at strong
coupling.
In this section we would like to analyze ways of counting BPS states in
string perturbation theory.

An important point that should be stressed from the beginning is the
following:
A generic BPS state $is$ $not$ protected from quantum corrections.
The reason is that sometimes groups of short BPS multiplets can combine
into
long multiplets of supersymmetry. Such long multiplets  are not
protected from non-renormalization theorems.
We would like thus to count BPS multiplicities in such a way that only
``unpaired" multiplets contribute.
As it is explained in Appendix B, this can be done with the help of
helicity supertrace formulae. They have precisely the properties we
need in order to count BPS multiplicities that are protected from
non-renormalization theorems.
Moreover, multiplicities counted via helicity supertraces are
insensitive to moduli.
They are the generalizations of the elliptic genus which is the stringy
generalization of the Dirac index.
In this sense, they are indices, insensitive to the details of the
physics.
We will show here how we can compute helicity supertraces in
perturbative string groundstates and we will work out some interesting
examples.

We will introduce the helicity generating partition
functions for $D=4$ string theories with $N\geq 1$ spacetime
supersymmetry.
The physical helicity in closed string theory $\l$ is a sum of the
left helicity $\l_L$ coming from the left movers and the right
helicity $\l_R$
coming from the right movers.
Then, we can consider the following helicity-generating partition
function
\be
Z(v,\bar v)=Str[q^{L_0}\;\bar q^{\bar L_0}e^{2\pi
iv\l_R-2\pi i\bar v\l_L}]
\,.\label{F1}\ee

We will first examine the heterotic string.
Four-dimensional vacua with at least N=1 spacetime supersymmetry
have
the following partition function
\be
Z^{\rm heterotic}_{D=4}={1\over \t_2\eta^2\bar \eta^2}\sum_{a,b=0}^1
{}~(-1)^{a+b+ab}~{\th[^a_b]\over \eta}~C^{Int}[^a_b]
\,,\label{F2}\ee
where we have separated the (light-cone) bosonic and fermionic
contributions of the four-dimensional  part.
$C[^a_b]$ is the partition function of the internal CFT with $(c,\bar
c)=(9,22)$ and at least (2,0) superconformal symmetry.
$a=0$ corresponds to the NS sector, $a=1$ to the R
sector
and $b=0,1$ indicates the presence of the projection $(-1)^{F_L}$,
where
$F_L$ is the zero mode of the N=2, U(1) current.

The oscillators that would contribute to the left helicity are the
left moving
light-cone bosons $\pd X^{\pm}=\pd X^3\pm i\pd X^4$ contributing
helicity $\pm 1$ respectively, and the the light-cone fermions
$\psi^{\pm}$ contributing again
$\pm 1$ to the left helicity.
Only $\bar\pd X^{\pm}$ contribute to the right-moving helicity.
Calculating (\ref{F1}) is straightforward with the result
\be
 Z^{\rm heterotic}_{D=4}(v,\bar v)={\xi(v)\bar\xi(\bar v)\over
\t_2\eta^2\bar \eta^2}\sum_{a,b=0}^1
{}~(-1)^{a+b+ab}~{\th[^a_b](v)\over \eta}~C^{Int}[^a_b]
\,,\label{F350}\ee
where $\xi(v)$ is given in (\ref{E19}).
This can be simplified using spacetime supersymmetry to
\be
 Z^{\rm heterotic}_{D=4}(v,\bar v)={\xi(v)\bar\xi(\bar v)\over
\t_2\eta^2\bar \eta^2}{\th[^1_1](v/2)\over \eta}~C^{Int}[^1_1](v/2)
\,,\label{F351}\ee
with
\be
C^{Int}[^1_1](v)=Tr_R[(-1)^{F^{Int}}~e^{2\pi i
v~J_0}~q^{L_0^{Int}-{3/8}}
{}~\bar q^{\bar L_0^{Int}-11/12}]
\,,\label{F4}\ee
where the trace is in the Ramond sector, and $J_0$ is the zero mode
of the U(1) current of the N=2 superconformal algebra.
$C^{Int}[^1_1](v)$ is the elliptic genus of the internal
(2,0)
theory and is antiholomorphic.
The leading term of $C^{Int}[^1_1](0)$ coincides with the Euler number
in CY
compactifications.

If we define
\be
Q={1\over 2\pi i}{\partial\over \partial v}\;\;\;,\;\;\;\bar
Q=-{1\over 2\pi i}{\partial\over \partial \bar v}
\,,\label{F5}\ee
then the helicity supertraces can be written as
\be
Str[\l^{2n}]=\left.(Q+\bar Q)^{2n}\;Z^{\rm heterotic}_{D=4}(v,\bar
v)\right|_{v=\bar v=0}
\,.\label{F6}\ee

Consider as an example the heterotic string on $T^6$ with
N=4, $D=4$ spacetime supersymmetry.
Its helicity partition function is
\be
Z^{\rm heterotic}_{N=4}(v,\bar v)={\vartheta^4_1(v/2)\over
\eta^{12}\bar \eta^{24}}\xi(v)\bar\xi(\bar v){\Gamma_{6,22}\over
\t_2}
\,.\label{F7}\ee

It is obvious that we need at least four powers of $Q$ in order to
get a non-vanishing contribution, implying $B_0=B_2=0$, in
agreement
with the N=4 supertrace formulae derived in Appendix B.
We will calculate $B_4$ which, according
to
(\ref{D26}),(\ref{D27}) is sensitive to short multiplets only:
\be
B_4=\langle (Q+\bar Q)^4\rangle=\langle
Q^4\rangle={3\over
2}{1\over \bar \eta^{24}}
\,.\label{F8}\ee
For the massless states the result agrees
with ({\ref{D27}), as it should.
Moreover, from (\ref{D26}) we observe that massive short multiplets
with a bosonic ground-state give an opposite contribution from
multiplets with a fermionic ground-state. We learn that all such
short massive multiplets  in
the heterotic spectrum
are "bosonic" with multiplicities given by the coefficients of the
$\eta^{-24}$.

Consider further
\be
B_6=\langle(Q+\bar Q)^6\rangle= \langle
Q^6+15Q^4\bar Q^2\rangle
={15\over 8}{2-\bar E_2\over \bar\eta^{24}}
\,.\label{F9}\ee
Since there can be no intermediate multiplets in the perturbative
heterotic spectrum we get only contributions from the short
multiplets. An explicit analysis at low levels confirms the agreement
between (\ref{D26}) and
(\ref{F9}).

For type-II vacua, there are fermionic contributions to the helicity
both from the left-moving and right-moving world-sheet fermions.
We will consider as a first  example the type-II string, compactified
on $T^6$
to four dimensions with maximal N=8 supersymmetry.

The light-cone helicity generating partition function is
\be
Z^{II}_{N=8}(v,\bar v)=Str[q^{L_0}\;\bar q^{\bar L_0}e^{2\pi
iv\l_R-2\pi i\bar v\l_L}]=
\label{cc2}\ee
$$={1\over
4}\sum_{\a,\b=0}^{1}\sum_{\bar\a,\bar\b=0}^{1}\;
(-1)^{\a+\b+\a\b}{\vartheta[^{\a}_{\b}](v)
\vartheta^3[^{\a}_{\b}](0)\over \eta^4}\;\;
(-1)^{\bar\a+\bar\b+\bar\a\bar
\b}{\bar\vartheta[^{\bar\a}_{\bar\b}]\bar\vartheta^3
[^{\bar\a}_{\bar\b}](0)\over \bar\eta^4}\;\;
{\xi(v)\bar\xi(\bar v)\over {\rm Im}\tau
|\eta|^4}\;\;{\Gamma_{6,6}\over
|\eta|^{12}}=
$$
$$={\Gamma_{6,6}\over {\rm Im}\tau }\;\;{\vartheta_1^4(v/2)\over
\eta^{12}}{\bar\vartheta_1^4(\bar v/2)\over
\bar\eta^{12}}\;\;\xi(v)\bar\xi(\bar v)\;\;.$$

It is obvious that in order to obtain a non-zero result, we need at
least a $Q^4$ on the left and a $\bar Q^4$ on the right.
This is in agreement with our statement in appendix B:
$B_0=B_2=B_4=B_6=0$ for an $N=8$ theory.
The first non-trivial case is $B_8$ and by straightforward
computation we obtain
\be
B_8\equiv Str[\l^8]=\langle(Q+\bar Q)^8\rangle =70\langle Q^4\bar
Q^4\rangle={315\over 2}{\Gamma_{6,6}\over {\rm Im}\tau }\;\;.
\label{cc6}\ee

At the massless level, the only N=8 representation is the supergravity
representation, which contributes $315/2$ in accordance with
(\ref{ee16}).
At the massive levels we have seen in appendix B that only short
representations
$S^j$ can contribute, each contributing $315/2\;(2j+1)$.
We learn from (\ref{cc6}) that all short massive multiplets
have $j=0$ and they are left and right ground states of the type II
CFT breaking thus N=8 supersymmetry to N=4.
Since the mass for these states is
\be
M^2={1\over 4}p_L^2\;\;\;,\;\;\;\vec m\cdot\vec n=0\;\;,
\label{cc7}\ee
such multiplets exist for any (6,6) lattice vector satisfying the
matching condition.
The multiplicity coming from the rest of the theory is one.

We will now compute the next non-trivial supertrace\footnote{We use
formulae from appendix C here.}
\be
B_{10}=\langle(Q+\bar Q)^{10}\rangle =210\langle Q^6\bar Q^4+Q^4\bar
Q^6\rangle=-{4725\over 8\pi^2}{\Gamma_{6,6}\over {\rm Im}\tau }\left(
{\vartheta_1'''\over \vartheta_1'}+3\xi''+\;cc\right)={4725\over
4}{\Gamma_{6,6}\over {\rm Im}\tau }\;.
\label{cc8}\ee

In this  trace, $I_1$ intermediate representations can also in
principle contribute. Comparing (\ref{cc8}) with
(\ref{ee13},\ref{ee23}) we learn that
there are no $I_1$ representations in the perturbative string
spectrum.

Moving further,
\be
B_{12}=\langle 495(Q^4\bar Q^8+Q^8\bar Q^4)+924Q^6\bar Q^6\rangle=
\left[{10395\over 2}+{31185\over 64}(E_4+\bar E_4)\right]
{\Gamma_{6,6}\over {\rm Im}\tau }
\label{cc9}\ee
$$=\left[{10395\cdot 19\over 32}+{10395\cdot 45\over
4}\left({E_4-1\over 240}+cc\right)\right]{\Gamma_{6,6}\over {\rm
Im}\tau }\;.
$$
Comparison with (\ref{ee19}) indicates that the first term in the
formula above contains the contribution of the short multiplets.
Here however, $I_2$ multiplets can also contribute and the second term
in (\ref{cc9}) describes precisely their contribution.
These are string states that are groundstates either on the left or
on the right and comparing with (\ref{ee28}) we learn that their
multiplicities are given by $(E_4-1)/240$.
More precisely, for a given mass level with $p_L^2-p_R^2=4N >0$
the multiplicity of these representations
at that mass level is given  by the sum of cubes of all divisors of
N, $d_4(N)$ (see Appendix C).
\be
I_2^j\;\;:\;\; \sum_j (-1)^{2j}D_j =d_4(N)\;\;.
\label{cc10}\ee
They break N=8 supersymmetry to N=2.

The last trace that long multiplets do not contribute is
\be
B_{14}=\langle (Q+\bar Q)^{14}\rangle=\left[{45045\over
32}20+{14189175\over
16}
\left(2{E_4-1\over 240}+{1-E_6\over 504}+cc\right)
\right]{\Gamma_{6,6}\over {\rm Im}\tau}\;\;.
\label{cc11}\ee
Although in this trace $I_3$ representations can contribute, there are
no such representations in the perturbative string spectrum.
The first term in (\ref{cc11}) comes from short representations while
the second
from $I_2$ representations.
Taking into account (\ref{ee29}) we can derive the following sum rule
\be
I_2^j\;\;:\;\; \sum_j (-1)^{2j}D_j^3 =d_6(N)\;\;.
\label{cc12}\ee

The final example we will consider is also instructive because it shows
that
although a string groundstate can contain many BPS multiplets, most of
them are not protected from renormalization.
The relevant vacuum is the type II string compactified on K3$\times
T^2$ down to four dimensions.

We will first start from the  $Z_{2}$ special point of the $K_3$
moduli space.
This is given by a  $Z_2$ orbifold of the four-torus. We can
write the one-loop vacuum amplitude as
\be
Z^{II}={1\over
8}\sum_{g,h=0}^{1}\sum_{\a,\b=0}^{1}\sum_{\bar\a,\bar\b=0}^{1}\;
(-1)^{\a+\b+\a\b}{\vartheta^2[^{\a}_{\b}]\over \eta^2}
{\vartheta[^{\a+h}_{\b+g}]\over \eta}
{\vartheta[^{\a-h}_{\b-g}]\over \eta}\times
\label{c1}\ee
$$\times(-1)^{\bar\a+\bar\b+\bar\a\bar \b}{\bar
\vartheta^2[^{\bar\a}_{\bar\b}]\over \bar\eta^2}
{\bar\vartheta[^{\bar\a+h}_{\bar\b+g}]\over \bar\eta}
{\bar\vartheta[^{\bar\a-h}_{\bar\b-g}]\over \bar\eta}\;\;
{1\over {\rm Im}\tau |\eta|^4}\;\;{\Gamma_{2,2}\over
|\eta|^4}\;\;Z_{4,4}[^h_g]$$
where
\be
Z_{4,4}[^0_0]={\Gamma_{4,4}\over
|\eta|^8}\;\;\;,\;\;\;Z_{4,4}[^0_1]=16{|\eta|^4\over
|\vartheta_2|^4}={|\vartheta_3\vartheta_4|^4\over |\eta|^8}
\label{c2a}\ee
\be
Z_{4,4}[^1_0]=16{|\eta|^4\over
|\vartheta_4|^4}={|\vartheta_2\vartheta_3|^4\over
|\eta|^8}\;\;\;,\;\;\;Z_{4,4}[^1_1]=16{|\eta|^4\over
|\vartheta_3|^4}={|\vartheta_2\vartheta_4|^4\over |\eta|^8}
\label{c2b}\ee

We have N=4 supersymmetry in four dimensions. The mass formula of BPS
states
depends only on the two-torus moduli.
Moreover states that are groundstates both on the left and the right
will give
short BPS multiplets that break half of the supersymmetry.
On the other hand states that are groundstates on the left but
otherwise
arbitrary on the right (and vice versa) will provide BPS states that
are intermediate multiplets breaking 3/4 of the supersymmetry.
Obviously there are many such states in the spectrum. Thus, we naively
expect
many perturbative intermediate multiplets.

We will now evaluate the helicity supertrace formulae.
We will first write the helicity generating function,
\be
Z^{II}(v,\bar v)={1\over
4}\sum_{\a\b\bar\a\bar\b}(-1)^{\a+\b+\a\b+\bar\a+\bar\b+\bar\a\bar\b}
{\vartheta[^{\a}_{\b}](v)\vartheta[^{\a}_{\b}](0) \over \eta^6}
{\bar\vartheta[^{\bar\a}_{\bar\b}](\bar v)\bar
\vartheta[^{\bar\a}_{\bar\b}](0)\over \bar\eta^6}\xi(v)\bar\xi(\bar
v)
C[^{\a\;\;\bar\a}_{\b\;\;\bar\b}]{\Gamma_{2,2}\over \tau_2}
\label{c33}\ee
$$
={\vartheta_1^2(v/2)\bar\vartheta^2_1(\bar v/2)\over
\eta^6\;\bar\eta^6}\xi(v)
\bar\xi(\bar v)C[^{1\;\;1}_{1\;\;1}](v/2,\bar v/2){\Gamma_{2,2}\over
\tau_2}
$$
where we have used the Jacobi
identity in the second line.
$C[^{\a\;\;\bar\a}_{\b\;\;\bar\b}]$ is the partition function of the
internal (4,4) superconformal field theory in the various sectors.
Moreover
$C[^{1\;\;1}_{1\;\;1}](v/2,\bar v/2)$ is an even function of $v,\bar
v$ due to the SU(2) symmetry and
\be
C[^{1\;\;1}_{1\;\;1}](v,0)=8\sum_{i=2}^4\;{\vartheta_i^2(v)\over
\vartheta_i^2(0)}
\label{c34}\ee
is the elliptic genus of the (4,4) internal theory on K3.
Although we calculated the elliptic genus in the $Z_2$ orbifold limit
the calculation is valid on the whole of K3 since the
elliptic genus does not depend on the moduli.

Let us first compute the trace of the fourth power of the helicity:
\be
\langle\l^4\rangle=\rangle (Q+\bar Q)^4\rangle =6\langle Q^2\bar Q^2
+Q^2\bar Q^4\rangle=36{\Gamma_{2,2}\over \tau_2}
\label{c355}\ee
As expected, we obtain contributions from the the groundstates only,
but with arbitrary  momentum and winding on the (2,2) lattice.
At the massless level, we have the N=4 supergravity multiplet
contributing 3
and 22 vector multiplets contributing 3/2 each, making a total of 36,
in agreement with (\ref{c355}).
There is a tower of massive short multiplets at each mass level, with
mass
$M^2=p_L^2$ where $p_L$ is the (2,2) momentum. The matching condition
implies, $\vec m\cdot \vec n=0$.

We will further  compute the trace of the sixth power of the helicity,
to investigate the presence of intermediate multiplets.
\be
\langle\l^6\rangle=\rangle (Q+\bar Q)^6\rangle =15\langle Q^4\bar Q^2
+Q^2\bar Q^4\rangle=90{\Gamma_{2,2}\over \tau_2}
\label{c35}\ee
where we have used
\be
\partial_v^2C[^{1\;\;1}_{1\;\;1}](v,0)|_{v=0}=
-16\pi^2\;E_2
\label{c36}\ee

The only contribution comes from the short multiplets again as
evidenced by
(\ref{D29}), since $22\cdot 15/8+13\cdot 15/4=90$.
We conclude that there are no contributions from intermediate
multiplets
in (\ref{c36}) although there are many such states in the spectrum.
The reason is that such intermediate multiplets pair up into long
multiplets.

We will finally comment on a problem where counting BPS multiplicities
is important.
This is the problem of counting black-hole microscopic states in the
case of maximal supersymmetry in type II string theory.
For an introduction we refer the reader to \cite{mal}.
The essential ingredient is that at weak coupling, states can be
constructed using various D-branes. At strong coupling these states
have the interpretation
of charged macroscopic black holes.
The number of states for given charges can be computed at weak
coupling.
These are BPS states. Their multiplicity can then be extrapolated to
strong coupling, and gives an entropy that scales as the classical area
of the black hole as postulated by Bekenstein and Hawking.
In view of our previous discussion such an extrapolation is naive.
It is the number of unpaired multiplets that can be extrapolated at
strong coupling.
Here however the relevant states are the lowest spin vector multiplets,
which
as shown in appendix B have always positive supertrace.
Thus, the total supertrace is proportional to the overall number of
multiplets
and justifies the naive extrapolation  to strong coupling.

\renewcommand{\theequation}{\thesection.\arabic{equation}}
\section{Outlook\label{out}}
\setcounter{equation}{0}

I hope to have provided a certain flavor of the the recent developments
towards a non-perturbative understanding of string theory.

Despite the many miraculous characteristics of string theory,
there are some major unresolved problems.
The most important in my opinion is to make contact with the real world
and more concretely to pin down the mechanism of supersymmetry breaking
and stability of the vacuum in that case.
Recent advances in our non-perturbative understanding of the theory
could help in this direction.

Also, the recent non-perturbative advances seem to require other
extended objects apart from strings.
This, makes the following question resurface: What is string theory?
A complete formulation which would include the extended objects
required is still lacking.

I think this  is an exciting period, because we seem being at the verge
to understand some of the mysteries of string theory.

\vskip 3cm

\addcontentsline{toc}{section}{Acknowledgments}
\section*{Acknowledgments}

I would like to thank the organizers of the school for their most warm
hospitality.

\vskip 3cm

\addcontentsline{toc}{subsection}{Appendix A: Toroidal Kaluza-Klein
reduction}
\section*{Appendix A: Toroidal Kaluza-Klein reduction}
\renewcommand{\theequation}{A.\arabic{equation}}
\setcounter{equation}{0}

In this appendix we will describe the Kaluza-Klein ans\"atz for
toroidal
dimensional reduction from 10 to $D<10$ dimensions.
A more detailed discussion can be found in \cite{ms}.
Hatted fields will denote the $(10-D)$-dimensional  fields and
similarly for the
indices.
Greek indices from the beginning of the alphabet will denote the
$10-D$
internal (compact) dimensions. Unhatted Greek indices from the middle
of
the alphabet will denote the $D$ non-compact dimensions.

The  standard form for the  10-bein is
\be
\hat e^{\hat r}_{\mh}=\left(\matrix{e^{r}_{\m}&
A^{\b}_{\m}E^{a}_{\b}\cr
0& E^{a}_{\a}\cr}\right)\;\;\;,\;\;\;\hat e^{\mh}_{\hat
r}=\left(\matrix{
e^{\m}_{r}& -e^{\n}_{r}A^{\a}_{\n}\cr
0&E^{\a}_{a}\cr}\right)
\,.\label{C1}\ee
For the metric we have
\be
\hat G_{\mh\nh}=\left(\matrix{g_{\m\n}+A_{\m}^{\a}G_{\a\b}A^{\b}_{\n}
& G_{\a\b}A^{\b}_{\m}\cr
G_{\a\b}A^{\b}_{\n}& G_{\a\b}\cr}\right)\;\;,\;\;\hat
G^{\mh\nh}=\left(\matrix{g^{\m\n}&-A^{\m\a}
\cr -A^{\n\a}&G^{\a\b}+A^{\a}_{\rho}A^{\b,\rho}\cr}\right)
\,.\label{C2}\ee
Then the part of the action containing the Hilbert term as well as
the dilaton
becomes
\be
\a'^{D-2}S_{D}^{heterotic}=\int d^D x
\sqrt{-{\rm det}~g}~e^{-\phi}\left[R+\pd_{\m}\phi\pd^{\m}\phi
+{1\over 4}\pd_{\m}G_{\a\b}\pd^{\m}G^{\a\b}-{1\over
4}G_{\a\b}{F^{A}_{\m\n}}^{\a}
F_A^{\b,\mu\nu}\right]
\,,\label{C3}\ee
where
\be
\phi=\hat \Phi-{1\over 2}\log ({\rm det} G_{\a\b})
\,,\label{C4}\ee
\be
{F^A_{\m\n}}^{\a}=\pd_{\m}A^{\a}_{\n}-\pd_{\n}A^{\a}_{\m}
\,.\label{C5}\ee
We will now turn to the antisymmetric tensor part of the action:
\be
-{1\over 12}\int d^{10} x\sqrt{-{\rm det}~ \hat G}e^{-\hat \Phi}\hat
H^{\mh\nh\rh}\hat H_{\mh\nh\rh}=-\int d^D x\sqrt{-{\rm det}~ g}~
e^{-\phi}\left[{1\over 4}H_{\m\a\b}H^{\m\a\b}+\right.
\label{C6}
\ee
$$\left.+{1\over 4}H_{\m\n\a}H^{\m\n\a}+{1\over
12}H_{\m\n\rho}H^{\m\n\rho}\right]
$$
where we have used $H_{\a\b\g}=0$, and
\be
H_{\m\a\b}=e^{r}_{\m}\hat e^{\mh}_{\hat r}\hat H_{\mh\a\b}=\hat
H_{\m\a\b}
\,,\label{C7}\ee
\be
H_{\m\n\a}=e^{r}_{\m}e^{s}_{\n}\hat e^{\mh}_{r}\hat e^{\nh}_{s}\hat
H_{\mh\nh\a}=\hat H_{\m\n\a}-A_{\m}^{\b}\hat
H_{\n\a\b}+A_{\n}^{\b}\hat H_{\m\a\b}
\,,\label{C8}\ee
\be
H_{\m\n\rho}= e^{r}_{\m}e^{s}_{\n}e^{t}_{\rho}\hat e^{\mh}_{r}\hat
e^{\nh}_{s}
\hat e^{\rh}_{t}\hat H_{\mh\nh\rh}=
\hat H_{\m\n\rho}+\left[-A_{\m}^{\a}\hat
H_{\a\n\rho}+A_{\m}^{\a}A_{\n}^{\b}\hat H_{\a\b\rho}+{\rm
cyclic}\right]
\,.\label{C9}\ee

Similarly,
\be
\int d^{10} x\sqrt{-{\rm det}~ \hat G}~e^{-\hat
\Phi}\sum_{I=1}^{16}\hat
F^{I}_{\mh\nh}F^{I,\mh\nh}=\int d^D x\sqrt{-{\rm det}~ g}~
e^{-\phi}\sum_{I=1}^{16}\left[\tilde F^{I}_{\m\n}\tilde
F^{I,\m\n}+2\tilde F^{I}_{\m\a}\tilde F^{I,\m\a}
\right]
\,,\label{C10}\ee
with
\be
Y^{I}_{\a}=\hat A^{I}_{\a}\;\;\;,\;\;\;A^{I}_{\m}=\hat
A^{I}_{\m}-Y^{I}_{\a}A^{a}_{\m}
\;\;\;,\;\;\;
\tilde F^{I}_{\m\n}=F^{I}_{\m\n}+Y^{I}_{\a}F^{A,\a}_{\m\n}
\label{C12}\ee
\be
\tilde F^{I}_{\m\a}=\pd_{\m}Y^{I}_{\a}\;\;\;,\;\;\;
F_{\m\n}^{I}=\pd_{\m}A^{I}_{\n}-\pd_{\n}A^{I}_{\m}
\,.\label{C14}\ee

We can now evaluate the D-dimensional antisymmetric tensor pieces
using (\ref{C7})-(\ref{C9}):

\be
\hat H_{\m\a\b}=\pd_{\m}\hat B_{\a\b}+{1\over
2}\sum_I\left[Y^{I}_{\a}\pd_{\m}
Y^{I}_{\b}-Y^{I}_{\b}\pd_{\m}Y^{I}_{\a}\right]
\,.\label{C15}\ee
Introducing
\be
C_{\a\b}\equiv\hat B_{\a\b}-{1\over 2}\sum_I Y^{I}_{\a}Y^{J}_{\b}
\,,\label{C16}\ee
we obtain from (\ref{C6})
\be
H_{\m\a\b}=\pd_{\m}C_{\a\b}+\sum_I Y^{I}_{\a}\pd_{\m}Y^{I}_{\b}
\,.\label{C166}\ee

Also
\be
\hat H_{\m\n\a}=
\pd_{\m}\hat B_{\n\a}-\pd_{\n}\hat B_{\m\a}+{1\over 2}\sum_I \left[
\hat A_{\n}^{I}\pd_{\m}Y_{\a}^{I}-\hat
A_{\m}^{I}\pd_{\n}Y^{I}_{\a}-Y^{I}_{\a}
\hat F^{I}_{\m\n}\right]
\,.\label{C17}\ee
Define
\be
B_{\m,\a}\equiv\hat B_{\m\a}+B_{\a\b}A_{\m}^{\b}+{1\over
2}\sum_I Y^{I}_{\a}A^{I}_{\m}
\,,\label{C18}\ee
\be
F^{B}_{\a,\m\n}=\pd_{\m}B_{\a,\n}-\pd_{\n}B_{\a,\m}
\,,\label{C19}\ee
we obtain from (\ref{C7})
\be
H_{\m\n\a}=F_{\a\m\n}^{B}-C_{\a\b}F^{A,\b}_{\m\n}-\sum_I Y^{I}_{\a}
F_{\m\n}^{I}
\,.\label{C20}\ee
Finally,
\be
B_{\m\n}=\hat B_{\m\n}+{1\over
2}\left[A^{\a}_{\m}B_{\n\a}+\sum_I A^{I}_{\m}
A^{\a}_{\n}Y^{I}_{\a}-(\m\leftrightarrow
\n)\right]-A^{\a}_{\m}A^{\b}_{\n}B_{\a\b}
\label{C21}\ee
and
\be
H_{\m\n\rho}=\pd_{\m}B_{\n\rho}-{1\over
2}\left[B_{\m\a}F^{A,\a}_{\n\rho}+A^{\a}_{\m}
F^{B}_{a,\n\rho}+\sum_I A^{I}_{\mu}F^{I}_{\n\rho}\right]+{\rm cyclic}
\label{C22}
\ee
$$\equiv \pd_{\m}B_{\n\rho}-{1\over 2}
L_{ij}A^{i}_{\m}F^{j}_{\n\rho}+{\rm cyclic}
$$
where we combined the $36-2D$ gauge fields
$A^{\a}_{\m},B_{\a,\mu},A^I_{\mu}$ into the uniform notation
$A^i_{\m}$, $i=1,2,\ldots,36-2D$ and
$L_{ij}$ is the O(10-D,26-D)-invariant metric.
We can combine the scalars $G_{\a\b},B_{\a\b},Y^I_{\a}$ into the
matrix $M$ given in (\ref{B5}).
Putting everything together,  the D-dimensional action becomes
\be
S^{\rm heterotic}_{D}=\int ~d^Dx~\sqrt{-{\rm
det}~g}e^{-\phi}\left[R+\pd^{\m}\phi\pd_{\m}\phi
-{1\over 12}\tilde H^{\m\n\rho}\tilde H_{\m\n\rho}-\right.
\label{C23}\ee
$$\left.-{1\over 4}(M^{-1})_{ij}
F^{i}_{\m\n}F^{j\m\n}+{1\over 8}Tr(\pd_{\m} M\pd^{\m}
M^{-1})\right]\;.
$$

We will also consider here the KK reduction of a three-index
antisymmetric tensor $C_{\m\n\rho}$. Such a tensor appears in type-II
string theory and eleven-dimensional supergravity.
The action for such a tensor is
\be
S_{C}=-{1\over 2\cdot 4!}\int d^d x\sqrt{-G}~\hat F^2\;\;,
\label{C24}\ee
where
\be
\hat F_{\m\n\rho\s}=\partial_{\mu}\hat C_{\n\rho\s}-\partial_{\s}\hat
C_{\m\n\rho}+
\partial_{\rho}
\hat C_{\s\m\n}-\partial_{\n}\hat C_{\rho\s\m}\;\;.
\label{C25}\ee

We define the lower-dimensional components as
\be
C_{\a\b\g}=\hat C_{\a\b\g}\;\;\;,\;\;\;C_{\m\a\b}=\hat
C_{\m\a\b}-C_{\a\b\g}A^{\g}_{\m}\;\;,
\label{C26}\ee
\be
C_{\m\n\a}=\hat C_{\m\n\a}+\hat C_{\m\a\b}A^{\b}_{\n}-\hat
C_{\n\a\b}A^{\b}_{\m}+C_{\a\b\g}A^{\b}_{\mu}A^{\g}_{\n}\;\;,
\label{C27}\ee
\be
C_{\m\n\rho}=\hat C_{\m\n\rho}+\left(-\hat
C_{\n\rho\a}A^{\a}_{\mu}+\hat C_{\a\b\rho}A^{\a}_{\m}A^{\b}_{\n}+{\rm
cyclic}\right)-C_{\a\b\g}
A^{\a}_{\m}A^{\b}_{\n}A^{\g}_{\rho}\;\;.
\label{C28}\ee
Then,
\be
S_{C}=-{1\over 2\cdot 4!}\int d^D x\sqrt{-g}\sqrt{{\rm
det}G_{\a\b}}\left[
F_{\m\n\rho\s}F^{\m\n\rho\s}+4F_{\m\n\rho\a}F^{\m\n\rho\a}+6
F_{\m\n\a\b}F^{\m\n\a\b}+4F_{\m\a\b\g}F^{\m\a\b\g}\right]\;,
\label{C29}\ee
where
\be
F_{\m\a\b\g}=\partial_{\m}C_{\a\b\g}\;\;\;,\;\;\;F_{\m\n\a\b}=
\partial_{\mu}C_{\n\a\b}-\partial_{\n}C_{\m\a\b}+
C_{\a\b\g}F^{\g}_{\m\n}\;,
\label{C30}\ee
\be
F_{\m\n\rho\a}=\partial_{\m}C_{\n\rho\a}+C_{\m\a\b}F^{\b}_{\n\rho}+
{\rm cyclic}\;\;,
\label{C31}\ee
\be
F_{\m\n\rho\s}=(\partial_{\m}C_{\n\rho\s}+{\rm
3~~perm.})+(C_{\rho\s\a}F^{\a}_{\m\n}+{\rm 5~~perm.})\;\;.
\label{C32}\ee

\addcontentsline{toc}{subsection}{Appendix B: BPS multiplets and
helicity supertrace formulae}
\section*{Appendix B: BPS multiplets and helicity supertrace
formulae\label{BPS}}
\renewcommand{\theequation}{B.\arabic{equation}}
\setcounter{equation}{0}

BPS states are important probes of non-perturbative physics in
theories with extended ($N\geq 2$) supersymmetry.

BPS states are special for the following reasons:

$\bullet$ Due to their relation with central charges, although
massive they form multiplets under extended supersymmetry which are
shorter than the generic massive multiplet.
Their mass is given in terms of their charges and moduli expectation
values.

$\bullet$ At generic points in moduli space  they are stable
due to energy and charge conservation.

$\bullet$ Their mass-formula is supposed to be exact if one uses
renormalized values for the charges and moduli.
\footnote{In theories with $N\geq 4$ supersymmetry there are no
renormalizations.}
The argument is that quantum corrections would spoil the relation of
mass and charges, and if we assume unbroken supersymmetry at the
quantum level
that would give incompatibilities  with the dimension of their
representations.

In order to present the concept of BPS states we will briefly review
the representation theory of $N$-extended supersymmetry.
A more complete treatment can be found in \cite{BW}.
The anti-commutation relations are
\be
\{Q_{\a}^I,Q_{\b}^J\}=\e_{\a\b}Z^{IJ}\;\;\;,\;\;\;
\{\bar Q_{\dot\a}^I,Q_{\dot\b}^J\}=\e_{\dot\a\dot\b}\bar Z^{IJ}
\;\;\;,\;\;\;\{Q_{\a}^I,\bar
Q_{\dot\a}^J\}=\d^{IJ}~2\s^{\mu}_{\a\dot\a}P_{\m}
\,,\label{3999}\ee
where $Z^{IJ}$ is the antisymmetric central charge matrix.

The algebra is invariant under the U(N) $R$-symmetry that rotates
$Q,\bar Q$.
We begin with a description of the representations of the algebra.
We will first assume that the central charges are zero.

$\bullet$ \underline{Massive representations}. We can go to the rest
frame
$P\sim(-M,\vec 0)$. The relations become
\be
\{Q_{\a}^I,\bar Q^J_{\dot\a}\}=2M\d_{\a\dot \a}\d^{IJ}
\;\;\;,\;\;\;\{Q^I_{\a},Q^J_{\b}\}=\{\bar Q^I_{\dot\a},\bar
Q^J_{\dot\b}\}
=0\,.\label{D1}\ee
Define the 2N fermionic harmonic creation and annihilation operators
\be
A^I_{\a}={1\over \sqrt{2M}}Q^I_{\a}\;\;\;,\;\;\;A^{\dagger
I}_{\a}={1\over \sqrt{2M}}\bar Q^I_{\dot\a}
\,.\label{D2}\ee
Building the representation is now easy. We start with Clifford
vacuum
$|\Omega\rangle$ which is annihilated by the $A^I_{\a}$ and we
generate
the representation by acting with the creation operators.
There are ${2N}\choose{n}$ states at the $n$-th oscillator level.
The total number of states is $\sum_{n=0}^{2N}$${2N}\choose{n}$, half
of them being bosonic
and half of them fermionic. The spin comes from symmetrization over
the spinorial indices. The maximal spin is the spin of the
ground-states plus $N$.

{\bf Example}. Suppose N=1 and the ground-state transforms into the
$[j]$ representation of SO(3).
Here we have two creation operators.
Then, the content of the massive representation is
$[j]\otimes([1/2]+2[0])
=[j\pm 1/2]+2[j]$.
The two spin-zero states correspond to the ground-state itself and to
the
state with two oscillators.

$\bullet$ \underline{Massless representations}. In this case we can
go to the frame $P\sim (-E,0,0,E)$.
The anti-commutation relations now become
\be
\{Q^I_{\a},\bar Q^{J}_{\dot\a}\}=2\left(\matrix{2E&0\cr
0&0\cr}\right)
\d^{IJ}
\,,\label{D3}\ee
the rest being zero. $Q_2^I,\bar Q_{\dot 2}^I$ totally anticommute so
they are represented by zero.
We have $N$ nontrivial creation and annihilation operators
$A^I=Q_1^I/2\sqrt{E}$,$A^{\dagger~I}=\bar Q_1^I/2\sqrt{E}$,
and the representation is $2^N$-dimensional.
It is much shorter than the massive one.

$\bullet$ \underline{Non-zero central charges}. In this case the
representations are massive. The central charge matrix
can be brought be a U(N) transformation to block diagonal form
and we will label the real positive eigenvalues by $Z_m$.
We assume that
$N$ is even so that $m=1,2,\ldots,N/2$.
We will split the index $I\to (a,m)$. $a=1,2$ labels positions inside
the $2\times 2$ blocks while $m$ labels the blocks.
Then
\be
\{Q_{\a}^{am},\bar Q_{\dot\a}^{bn}\}=2M\d^{\a\dot\a}\d^{ab}\d^{mn}
\;\;\;,\;\;\;\{Q_{\a}^{am},Q_{\b}^{bn}\}=Z_n\e^{\a\b}\e^{ab}\d^{mn}
\,.\label{D4}\ee
Define the following fermionic oscillators
\be
A^{m}_{\a}={1\over
\sqrt{2}}[Q^{1m}_{\a}+\e_{\a\b}Q^{2m}_{\b}]\;\;\;,\;\;\;
B^{m}_{\a}={1\over \sqrt{2}}[Q^{1m}_{\a}-\e_{\a\b}Q^{2m}_{\b}]
\,,\label{D5}\ee
and similarly for the conjugate operators.
The anticommutators become
\be
\{A^m_{\a},A^n_{\b}\}=\{A^m_{\a},B^n_{\b}\}=\{B^m_{\a},B^n_{\b}\}=0
\,,\label{D6}\ee
\be
\{A^{m}_{\a},A^{\dagger
n}_{\b}\}=\d_{\a\b}\d^{mn}(2M+Z_n)\;\;\;,\;\;\;
\{B^{m}_{\a},B^{\dagger n}_{\b}\}=\d_{\a\b}\d^{mn}(2M-Z_n)
\,.\label{D7}\ee
Unitarity requires that the right hand sides in (\ref{D7}) be
non-negative.
This in turn implies the Bogomolnyi bound
\be
M\geq {\rm max}\left[ {Z_{n}\over 2}\right]
\,.\label{D8}\ee
Consider $0\leq r\leq N/2$ of the $Z_n$'s to be equal to $2M$.
Then $2r$ of the $B$-oscillators vanish identically and we are left
with $2N-2r$ creation and annihilation operators.
The representation has $2^{2N-2r}$ states.
The maximal case $r=N/2$ gives rise to the short BPS multiplet whose
number of states are the same as in the massless multiplet.
The other multiplets with $0<r<N/2$ are known as intermediate BPS
multiplets.

Another ingredient that makes supersymmetry special is some special
properties of supertraces of powers of the helicity.
Such supertraces appear in loop amplitudes and they will be quite
useful.
They can also be used to distiguish BPS states.
We will define the helicity supertrace on a supersymmetry
representation $R$
as
\be
B_{2n}(R)=Tr_R[(-1)^{2\l}\l^{2n}]
\,.\label{D9}\ee
It is useful to introduce the ``helicity generating function" of a
given supermultiplet R
\be
Z_{R}(y) = str\  y^{2 \lambda}
\,.\label{D10}\ee
For a particle of spin $j$ we have
\be
Z_{[j]} = \cases{& $(-)^{2j} \Bigl( { y^{2j+1}-y^{-2j-1}
\over y- 1/y } \Bigr)$ \ \  {\rm massive} \cr
 &\ \cr
 & $(-)^{2j} ( y^{2j}+y^{-2j})$ \ \ \ \
  {\rm massless} \cr}
\,.\label{D11}
\ee
When tensoring representations the generating functionals get
multiplied,
\be
Z_{r\otimes {\tilde r}} = Z_r Z_{\tilde r}
\,.\label{D12}\ee
The supertrace of the $n$th power of helicity can be extracted
from the generating functional through
\be
B_n(R) =   (y^2{d\over dy^2})^n \ Z_R(y)\vert_{y=1}
\,.\label{D13}\ee

For a supersymmetry representation constructed from a spin $[j]$
ground-state
by acting with $2m$ oscillators we obtain
\be
Z_{m}(y)=Z_{[j]}(y)(1-y)^m(1-1/y)^m
\,.\label{D14}\ee

We will now analyse in more detail N=2,4 supersymmetric
representations

$\bullet$ \underline{N=2 Supersymmetry}. There is only one central
charge eigenvalue $Z$.
The long massive representations has the following content:
\be
L_j\;\;\;:\;\;\;[j]\otimes([1]+4[1/2]+5[0])
\,.\label{D33}\ee

When $M=Z/2$ we obtain the short (BPS) massive multiplet
\be
S_j\;\;\;:\;\;\;[j]\otimes(2[1/2]+4[0])
\,.\label{D34}\ee

Finally the massless multiplets have the following content
\be
M^0_{\l}\;\;\;:\;\;\;\pm (\l+1/2)+2(\pm\l)+\pm(\l-1/2)
\,.\label{D35}\ee
$\l=0$ corresponds to the hypermultiplet, $\l=1/2$ to the vector
multiplet
and $\l=3/2$ to the supergravity multiplet.

We have the following helicity supertraces
\be
B_{0}({\rm any~~rep})=0
\,,\label{D36}\ee

\be
B_2(M^0_{\l})=(-1)^{2\l+1}\;\;\;,\;\;\;B_2(S_j)=
(-1)^{2j+1}~D_j\;\;\;,\;\;\;
B_2(L_j)=0
\,.\label{D37}\ee

$\bullet$ \underline{N=4 Supersymmetry}.  Here we have two eigenvalues
for the central charge matrix $Z_1\geq Z_2\geq 0$.
For the  generic massive multiplet, $M>Z_1$,
and all eight  raising operators
act non-trivially.  The representation is long,
 containing 128 bosonic
and 128 fermionic states.
The generic long massive multiplet can be generated by tensoring the
representation $[j]$ of its ground-state with the long fermionic
oscillator
representation of the N=4 algebra:

\be
{\rm L_j}\;\;:\;\; [j]\otimes
\left(42[0]+48[1/2]+27[1]+8[3/2]+[2]\right)
\,.\label{D15}\ee
It contains $128\;D_j$ bosonic degrees of freedom and $128\;D_j$
fermionic
ones ($D_j=2j+1$).
The  minimum-spin
massive long (ML) multiplet has $j=0$ and maximum spin 2  with the
following content:
\be
s=2\;\;{ massive}\;\;{ long}\;:\;\;42[0]+48[1/2]+27[1]+8[3/2]+[2]
\,.\label{a1}\ee

The generic  representation saturating the mass bound,
 $M=Z_{1}>Z_{2}$,  leaves  one unbroken  supersymmetry
 and is referred to as  massive intermediate BPS multiplet .
It can be obtained as
\be
I_j\;\;:\;\;[j]\otimes(14[0]+14[1/2]+6[1]+[3/2])
\label{D16}\ee
and contains 32$D_j$ bosonic and $32D_j$ fermionic states.
The minimum spin multiplet (j=0) has maximum spin 3/2 and content

\be
I_{3/2}\;\;:\;\;14[0]+14[1/2]+6[1]+[3/2]
\,.\label{D17}\ee

Finally, when $M=|Z_1|=|Z_2|$ the representation is a short
BPS representation.
It breaks half of the supersymmetries. For massive such
representations
we have the content
\be
S_j\;\;:\;\;[j]\otimes(5[0]+4[1/2]+[1])
\,,\label{D18}\ee
with $8D_j$ bosonic and $8D_j$ fermionic states.
The representation with minimum greatest spin is the one with $j=0$,
and
maximum spin 1:
\be
S_1\;\;:\;\;5[0]+4[1/2]+[1]
\,.\label{D19}\ee

Massless multiplets, which arise only when both  central charges
vanish, are thus
always short.
They have the following O(2) helicity content:
\be
M^0_{\l}\;\;:\;\;[\pm (\l+1)]+4[\pm(\l+1/2)]+6[\pm (\l)]+
4[\pm(\l-1/2)]+[\pm(\l-1)]
\,,\label{D20}\ee
with 16 bosonic and 16 fermionic states.
There is also the CPT-self-conjugate vector representation ($V^0$)
(corresponding to $\l=0$)
with content $6[0]+4[\pm 1/2]+[\pm 1]$ and 8 bosonic and 8 fermionic
states.
For $\l=1$ we obtain  the spin-two massless supergravity
multiplet which has the  helicity content
\be
M^0_1\;\;\;:\;\;[\pm 2]+4[\pm 3/2]+6[\pm 1]+4[\pm 1/2]+2[0]
\,.\label{D21}\ee
Long representations can be decomposed
into intermediate representations as
\be
L_{j}\to 2\;I_{j}+I_{j+1/2}+I_{j-1/2}
\,.\label{D22}\ee
When further, by varying the moduli, we can arrange that
$M=|Z_1|=|Z_2|$
then the massive intermediate representations can break into massive
short
representations as
\be
I_j\to 2S_j+S_{j+1/2}+S_{j-1/2}
\,.\label{D23}\ee
Finally when a short representation becomes massless, it decomposes
as follows into massless representations:
\be
S_j\to \sum_{\lambda=0}^{j}
M^0_{\lambda}
\;\;\;,\;\;  j-\l \in \Z
\,.\label{D24}\ee

By direct calculation we obtain the following helicity supertrace
formulae:

\be
B_n({\rm any\;\;rep})=0\;\;\;{\rm for}\;\;\;n=0,2
\,.\label{D25}\ee
The non-renormalization of the two derivative effective actions in
N=4 supersymmetry is based on (\ref{D25}).

\be
B_4(L_j)=B_4(I_j)=0\;\;\;,\;\;\;B_4(S_j)=(-1)^{2j}{3\over
2}D_j\label{D26}\ee
\be
B_4(M^0_{\lambda})=(-1)^{2\lambda}\;3\;
\;\;,\;\;\;B_4(V^0)={3\over 2}
\,.\label{D27}\ee
These imply that only short multiplets contribute in the
renormalization
of some terms in the four derivative effective action in the presence
of N=4
supersymmetry.
It also strongly suggests that such corrections come only from one
order (usually one-loop) in perturbation theory.

The following helicity sums will be useful when counting intermediate
multiplets in string theory:
\be
B_6(L_j)=0\;\;\;,\;\;\;B_6(I_j)=(-1)^{2j+1}{45\over
4}D_j\;\;\;,\;\;\;B_6(S_j)=(-1)^{2j}{15\over 8}D^3_j
\,,\label{D28}\ee
\be
B_6(M^0_{\lambda})=(-1)^{2\lambda}{15\over 4}(1+12\lambda^2)\;
\;\;,\;\;\;B_6(V^0)={15\over 8}
\,.\label{D29}\ee

Finally,
\be
B_8(L_j)=(-1)^{2j}{315\over
4}D_j\;\;\;,\;\;\;B_8(I_j)=(-1)^{2j+1}{105\over 16}D_j(1+D^2_j)
\,,\label{D30}\ee
\be
B_8(S_j)=(-1)^{2j}{21\over 64}D_j(1+2\;D^4_j)
\,,\label{D31}\ee
\be
B_8(M^0_{\lambda})=(-1)^{2\lambda}{21\over
16}(1+80\lambda^2+160\lambda^4)\;
\;\;,\;\;\;B_8(V^0)={63\over 32}
\,.\label{D32}\ee
The massive long N=4 representation is the same as the short massive
N=8 representation, which explains the result in (\ref{D30}).

Observe that the trace formulae above are in accord with the
decompositions
(\ref{D22})-(\ref{D24}).

\vskip .5cm

$\bullet$\underline{N=8 supersymmetry}.
The highest possible supersymmetry in four dimensions is N=8.
Massless representations ($T_0^{\lambda}$), have the following
helicity content
\be
(\l\pm 2)+8\left(\l\pm {3\over 2}\right)+28(\l\pm 1)+56\left(\l\pm
{1\over
2}\right)
+70(\l)\;\;.
\label{e1}\ee
Physical (CPT-invariant) representations are given by
$M_0^{\l}=T_0^{\l}+T_0^{-\l}$
and contain $2^8$ bosonic states and an equal number of fermionic ones
with the exception of the supergravity representation $M_0^0=T_0^0$
which is CPT-self-conjugate:
\be
(\pm 2)+8\left(\pm {3\over 2}\right)+28(\pm 1)+56\left(\pm {1\over
2}\right)+70(0)\;\;,
\label{e2}
\ee
and contains $2^7$ bosonic states.

Massive short representations ($S^j$), are labeled by the SU(2) spin j
of the ground state and have the following content
\be
[j]\otimes \left([2]+8[3/2]+27[1]+48[1/2]+42[0]\right)\;\;.
\label{e3}
\ee
They break four (half) of the supersymmetries and contain $2^7\cdot
D_j$ bosonic states.
$S^j$ decomposes to massless representations as
\be
S^j\to \sum_{\l=0}^j\;M_0^{\l}\;\;,
\label{e222}\ee
where the sum runs on integer values of $\l$ if $j$ is integer and on
half-integer values if $j$ is half-integer.

There are three types of intermediate multiplets which we list below
\be
I_1^j\;:\;[j]\otimes
\left([5/2]+10[2]+44[3/2]+110[1]+165[1/2]+132[0]\right)\;\;,
\label{e4}
\ee
\be
I_2^j\;:\;[j]\otimes
\left([3]+12[5/2]+65[2]+208[3/2]+429[1]+572[1/2]+429[0]\right)\;\;,
\label{e5}
\ee
\be
I_3^j\;:\;[j]\otimes
\left([7/2]+14[3]+90[5/2]+350[2]+910[3/2]+1638[1]+
2002[1/2]+1430[0]\right)\;.
\label{e6}
\ee
They break respectively 5,6,7 supersymmetries.
They contain $2^9\cdot D_j$ ($I_1^j$), $2^{11}\cdot D_j$ ($I_2^j$)
and
$2^{13}\cdot D_j$ ($I_3^j$) bosonic states.

Finally, the long representations ($L^j$) (that break all
supersymmetries )
are given by
\be
[j]\otimes
\left([4]+16[7/2]+119[3]+544[5/2]+1700[2]+3808[3/2]+
6188[1]+7072[1/2]+4862
[0]\right).
\label{e7}
\ee
$L^j$ contains $2^{15}\cdot D_j$ bosonic states.

We also have the following recursive decomposition formulae:
\be
L^j\to I_3^{j+{1\over 2}}+2 I_3^j+I_3^{j-{1\over 2}}\;\;,
\label{e8}
\ee
\be
I_3^j\to I_2^{j+{1\over 2}}+2 I_2^j+I_2^{j-{1\over 2}}\;\;,
\label{e9}
\ee
\be
I_2^j\to I_1^{j+{1\over 2}}+2 I_1^j+I_1^{j-{1\over 2}}\;\;,
\label{e10}
\ee
\be
I_1^j\to S^{j+{1\over 2}}+2 S^j+S^{j-{1\over 2}}\;\;.
\label{e11}
\ee

All even helicity supertraces up to order six vanish for N=8
representations.
For the rest we obtain:
\be
B_8(M_0^{\l})=(-1)^{2\l}\;315\;\;,
\label{ee12}\ee
\be
B_{10}(M_0^{\l})=(-1)^{2\l}\;{4725\over 2}(6\l^2+1)\;\;,
\label{ee13}\ee
\be
B_{12}(M_0^{\l})=(-1)^{2\l}\;{10395\over 16}(240\l^4+240\l^2+19)\;\;,
\label{ee14}\ee
\be
B_{14}(M_0^{\l})=(-1)^{2\l}\;{45045\over
16}(336\l^6+840\l^4+399\l^2+20)\;\;,
\label{ee15}\ee
\be
B_{16}(M_0^{\l})=(-1)^{2\l}\;{135135\over 256}(7680\l^8+35840\l^6
+42560\l^4+12800\l^2+457)\;\;,
\label{ee16}\ee

The supertraces of the massless supergravity representation $M_0^0$
can be obtained from the above by setting $\l=0$ and dividing by a
factor of two to account for the smaller dimension of the
representation.

\be
B_8(S^j)=(-1)^{2j}\cdot {315\over 2}D_j\;\;,
\label{ee17}\ee
\be
B_{10}(S^j)=(-1)^{2j}\cdot {4725\over 8}D_j(D_j^2+1)\;\;,
\label{ee18}\ee
\be
B_{12}(S^j)=(-1)^{2j}\cdot {10395\over 32}D_j(3D_j^4+10D_j^2+6)\;\;,
\label{ee19}\ee
\be
B_{14}(S^j)=(-1)^{2j}\cdot {45045\over
128}D_j(3D_j^6+21D_j^4+42D_j^2+14)\;\;,
\label{ee20}\ee
\be
B_{16}(S^j)=(-1)^{2j}\cdot {45045\over
512}D_j(10D_j^{8}+120D_j^6+504D_j^4+560D_j^2+177)\;\;,
\label{ee21}\ee

\be
B_8(I_1^j)=0\;\;,
\label{ee22}\ee
\be
B_{10}(I_1^j)=(-1)^{2j+1}\cdot {14175\over 4}D_j\;\;,
\label{ee23}\ee
\be
B_{12}(I_1^j)=(-1)^{2j+1}\cdot {155925\over 16}D_j(2D_j^2+3)\;\;,
\label{ee24}\ee
\be
B_{14}(I_1^j)=(-1)^{2j+1}\cdot  {2837835\over
64}D_j(D_j^2+1)(D_j^2+4)\;\;,
\label{ee25}\ee
\be
B_{16}(I_1^j)=(-1)^{2j+1}\cdot {2027025\over
128}D_j(4D_j^6+42D_j^4+112D_j^2+57)\;\;,
\label{ee26}\ee

\be
B_8(I_2^j)=B_{10}(I_2^j)=0\;\;,
\label{ee27}\ee
\be
B_{12}(I_2^j)=(-1)^{2j}\cdot {467775\over 4}D_j\;\;,
\label{ee28}\ee
\be
B_{14}(I_2^j)=(-1)^{2j}\cdot {14189175\over 16}D_j(D_j^2+2)\;\;,
\label{ee29}\ee
\be
B_{16}(I_2^j)=(-1)^{2j}\cdot {14189175\over
32}D_j(6D_j^4+40D_j^2+41)\;\;,
\label{ee30}\ee

\be
B_8(I_3^j)=B_{10}(I_3^j)=B_{12}(I_3^j)=0\;\;,
\label{ee31}\ee
\be
B_{14}(I_3^j)=(-1)^{2j+1}\cdot {42567525\over 8}D_j\;\;,
\label{ee32}\ee
\be
B_{16}(I_3^j)=(-1)^{2j+1}\cdot {212837625\over 8}D_j(2D_j^2+5)\;\;,
\label{ee33}\ee

\be
B_8(L^j)=B_{10}(L^j)=B_{12}(L^j)=B_{14}(L^j)=0\;\;,
\label{ee34}\ee
\be
B_{16}(L^j)=(-1)^{2j}\cdot {638512875\over 2}D_j\;\;.
\label{ee35}\ee

A further check of the formulae above is provided by the fact that
they
respect the decomposition formulae of the various representations
, (\ref{e222},\ref{e8}-\ref{e11}).

\addcontentsline{toc}{subsection}{Appendix C: Modular forms}
\section*{Appendix C: Modular forms}
\renewcommand{\theequation}{C.\arabic{equation}}
\setcounter{equation}{0}

In this appendix we collect some formulae
for modular forms, which are useful for analysing the spectrum of BPS
states and
BPS-generated one-loop corrections to the effective supergravity
theories.
A (holomorphic) modular form $F_{d}(\tau)$ of weight $d$ behaves as
follows under modular transformations:
\be
F_d(-1/\t)=\t^{d}F_d(\t)\;\;\;F_d(\t+1)=F_d(\t)
\,.\label{E1}\ee

We list first the  Eisenstein series:
\be
E_2={12\over i\pi}\partial_{\tau}\log\eta=1-24\sum_{n=1}^{\infty}
{nq^n\over 1-q^n}
\,,\label{E2}\ee
\be
E_4={1\over 2}\left(\vartheta_2^8+\vartheta_3^8+\vartheta_4^8\right)
=1+240\sum_{n=1}^{\infty}{n^3q^n\over 1-q^n}
\,,\label{E3}\ee
\be
E_6={1\over 2}\left(\vartheta_2^4+\vartheta_3^4\right)
\left(\vartheta_3^4+\vartheta_4^4\right)
\left(\vartheta_4^4-\vartheta_2^4\right)=
1-504\sum_{n=1}^{\infty}{n^5q^n\over 1-q^n}
\,.\label{E4}\ee
In counting  BPS states in string theory the following combinations
arise
\be
H_2\equiv {1-E_2\over 24}=\sum_{n=1}^{\infty}{nq^n\over 1-q^n}\equiv
\sum_{n=1}^{\infty}d_2(n)q^n
\,,\label{E5}\ee
\be
H_4\equiv {E_4-1\over 240}=\sum_{n=1}^{\infty}{n^3q^n\over
1-q^n}\equiv
\sum_{n=1}^{\infty}d_4(n)q^n
\,,\label{E6}\ee
\be
H_6\equiv {1-E_6\over 504}=\sum_{n=1}^{\infty}{n^5q^n\over
1-q^n}\equiv
\sum_{n=1}^{\infty}d_6(n)q^n
\,.\label{E7}\ee
We have the following arithmetic formulae for $d_{2k}$:
\be
d_{2k}(N)=\sum_{n|N}\;n^{2k-1}\;\;\;,\;\;\;k=1,2,3
\,.\label{E8}\ee

$E_4$ and $E_6$ are modular forms of weight four and six
respectively.
They generate the ring of modular forms.
$E_2$ is not exactly a modular form.
However,
\be
\hat E_2=E_2-{3\over \pi\t_2}
\label{E9}\ee
is a modular form of weight two but is not holomorphic any more.
The (modular invariant) $j$ function and $\eta^{24}$ can be written
as
\be
j={E_4^3\over \eta^{24}}={1\over q}+744+\ldots\;\;\;,\;\;\;
\eta^{24}={1\over 2^6\cdot 3^3}\left[E_4^3-E_6^2\right]
\,.\label{E10}\ee

Here we will give some identities between derivatives of
$\th$-functions and
modular forms. They are useful for trace computations in
string theory.
 \be
{\vartheta_{1}'''\over \vartheta_1'}=-\pi^2\;E_2
\;\;\;,\;\;\;
{\vartheta^{(5)}_1\over \vartheta_1'}=-\pi^2\;E_2\left(4\pi
i\partial_{\tau}
\log E_2-\pi^2 E_2\right)
\,,\label{E16}\ee
\be
-3{\vartheta^{(5)}_1\over \vartheta_1'}+5\left({\vartheta_1'''\over
\vartheta_1'}\right)^2=2\pi^4 E_4
\,,\label{E17}\ee
\be
-15{\vartheta^{(7)}_1\over \vartheta_1'}-{350\over
3}\left({\vartheta_1'''\over
\vartheta_1'}\right)^3+105{\vartheta_1^{(5)}\vartheta_1'''\over
\vartheta_1'^2}={80\pi^6\over 3}E_6
\,,\label{E18}\ee
\be
{1\over 2}\sum_{i=2}^4~{\th_i''~\th_i^7\over (2\pi i)^2}={1\over 12}
(E_2E_4-E_6)
\,.\label{E188}\ee

The function $\xi(v)$ that appears in string helicity generating
partition functions is defined as
\be
\xi(v)=\prod_{n=1}^{\infty}{(1-q^n)^2\over (1-q^ne^{2\pi iv})
(1-q^ne^{-2\pi iv})}={\sin\pi v\over \pi}{\vartheta_1'\over
\vartheta_1(v)}\;\;\;\,\;\;\;\xi(v)=\xi(-v)
\,.\label{E19}\ee
It satisfies
\be
\xi(0)=1
\;\;\;,\;\;\;
\xi^{(2)}(0)=-{1\over 3}\left(\pi^2+{\vartheta_1'''\over
\vartheta_1'}\right)=-{\pi^2\over 3}(1-E_2)
\,,\label{E21}\ee
\be
\xi^{(4)}(0)={\pi^4\over 5}+{2\pi^2\over 3}{\vartheta_1'''\over
\vartheta_1'}
+{2\over 3}\left({\vartheta_1'''\over \vartheta_1'}\right)^2-{1\over
5}{\vartheta^{(5)}_1\over \vartheta_1'}={\pi^4\over
15}(3-10E_2+2E_4+5E_2^2)
\,,\label{E22}\ee

\be
\xi^{(6)}(0)=-{\pi^6\over 7}-\pi^4{\vartheta_1'''\over \vartheta_1'}
-{10\pi^2\over 3}\left({\vartheta_1'''\over
\vartheta_1'}\right)^2+\pi^2
{\vartheta^{(5)}_1\over \vartheta_1'}-{10\over 3}
\left({\vartheta_1'''\over
\vartheta_1'}\right)^3+2{\vartheta_1^{(5)}\vartheta_1'''\over
\vartheta_1'^2}
-{1\over 7}{\vartheta^{(7)}_1\over \vartheta_1'}=
\label{E23}\ee
$$={\pi^6\over 63}(-9+63E_2-105E_2^2-42E_4+16E_6+42E_2E_4+35E_2^3)
$$
where $\xi^{(n)}(0)$ stands for taking the $n$-th derivative with
respect
to $v$ and then setting $v=0$.

\addcontentsline{toc}{subsection}{Appendix D: Electric-Magnetic duality
in D=4}
\section*{Appendix D: Electric-Magnetic duality in D=4}
\renewcommand{\theequation}{D.\arabic{equation}}
\setcounter{equation}{0}

In this appendix we will describe electric-magnetic duality
transformations
for free gauge fields.
We consider here a collection of abelian gauge fields in $D=4$.
In the presence of supersymmetry  we can write terms quadratic in the
gauge
fields as

\be
L_{gauge}=-{1\over 8}{\rm Im}\int d^4 x\sqrt{-{\rm det}g}\; {\bf
F}^{i}_{\m\n}N_{ij}{\bf F}^{j,\m\n}
\,,\label{H1}\ee
where
\be
{\bf F}_{\m\n}=F_{\m\n}+i^*F_{\m\n}\;\;\;,\;\;\;
^*F_{\m\n}={1\over 2}{{\e_{\m\n}}^{\rho\s}\over \sqrt{-
g}}F_{\rho\s}
\,,\label{H2}\ee
with the property (in Minkowski space) that $^{**}F=-F$ and
$^*F_{\m\n}\;  ^*F^{\m\n}=-F_{\m\n}F^{\m\n}$.

In components,  the langrangian (\ref{H1}) becomes
\be
L_{\rm gauge}=-{1\over 4}\int d^4x
\left[\sqrt{-g}\;F^{i}_{\m\n}N^{ij}_2\;F^{j,\m\n}+F^{i}_{\m\n}
N_1^{ij})\;^*F^{j,\m\n}\right]
\,.\label{H3}\ee

Define now the tensor that gives the equations of motion
\be
{\bf G}^{i}_{\m\n}=N_{ij}{\bf F}^{j}_{\m\n}=N_1\;F-N_2 \;^*F+i
(N_2\;F+N_1\;^*F)
\,,\label{H4}\ee
with $N=N_1+iN_2$.
The equations of motion can be written in the form
${\rm Im}\na^{\m}{\bf G}^{i}_{\m\n}=0$,
while the Bianchi identity is
${\rm Im}\na^{\m}{\bf F}^{i}_{\m\n}=0$,
or
\be
{\rm Im}\na^{\m}\left(\matrix{{\bf G}^{i}_{\m\n}\cr {\bf
F}^{i}_{\m\n}\cr}\right)=\left(\matrix{0\cr 0\cr}\right)
\,.\label{H5}\ee
Obviously any Sp(2r,R) transformation of the form
\be
\left(\matrix{{\bf G'}_{\m\n}\cr {\bf F'}_{\m\n}\cr}\right)=
\left(\matrix{A&B\cr C&D\cr}\right)\left(\matrix{{\bf G}_{\m\n}\cr
{\bf F}_{\m\n}\cr}\right)\,,\label{H6}\ee
where $A,B,C,D$ are $r\times r$ matrices ($CA^{t}-AC^{t}=0$,
$B^{t}D-D^{t}B=0$, $A^{t}D-C^{t}B={\bf 1}$),
preserves the collection of equations of motion and Bianchi
identities.
At the same time
\be
N'=(AN+B)(CN+D)^{-1}
\,.\label{H7}\ee

The duality transformations are
\be
F'=C(N_1\;F-N_2
\;^*F)+D\;F\;\;\;,\;\;\;^*F'=C(N_2\;F+N_1\;^*F)+D\;^*F
\,.\label{H8}\ee
In the simple case $A=D={\bf 0}$, $-B=C={\bf 1}$ they become
\be
F'=N_1\;F-N_2
\;^*F\;\;\;,\;\;\;^*F'=N_2\;F+N_1\;^*F\;\;\;,\;\;\;N'=-{1\over N}
\,.\label{H9}\ee
When we perform duality with respect to one of the gauge fields (we
will call its component $0$) we have
\be
\left(\matrix{A&B\cr C&D\cr}\right)=\left(\matrix{{\bf 1}-e&-e\cr
e&{\bf 1}-e\cr}\right)\;\;\;,\;\;\;e=\left(\matrix{1&0&...\cr
0&0&...\cr
.&.&\cr}\right)
\,.\label{H10}\ee
\be
N_{00}'=-{1\over N_{00}}\;\;,\;\;N'_{0i}={N_{0i}\over
N_{00}}\;\;,\;\;
N_{i0}'={N_{i0}\over
N_{00}}\;\;,\;\;N_{ij}'=N_{ij}-{N_{i0}N_{0j}\over N_{00}}
\,.\label{H11}\ee

Finally consider the duality generated by
\be
\left(\matrix{A&B\cr C&D\cr}\right)=\left(\matrix{{\bf 1}-e_1&e_2\cr
-e_2&{\bf
1}-e_1\cr}\right)\;\;\;,\;\;\;e_1=\left(\matrix{1&0&0&...\cr
0&1&0&...\cr
0&0&0&.\cr .&.&.&.\cr}\right)\;\;,\;\;e_2=\left(\matrix{0&1&0&...\cr
-1&0&0&...\cr
0&0&0&.\cr .&.&.&.\cr}\right)
\,.\label{H13}\ee
We will denote the indices in the 2-d subsector where the duality
acts by $\a,\b,\g ...$.
Then
\be
N'_{\a\b}=-{N_{\a\b}\over {\rm det}N_{\a\b}}\;\;\;N'_{\a
i}=-{N_{\a\b}\e^{\b\g}N_{\g i}\over {\rm det}N_{\a\b}}\;\;\;,\;\;\;
N'_{i\a}={N_{i\b}\e^{\b\g}N_{\a\g}\over {\rm det}N_{\a\b}}
\,,\label{H14a}\ee
\be
N'_{ij}=N_{ij}+{N_{i\a}\e^{\a\b}N_{\b\g}\e^{\g\d}N_{\d j}\over {\rm
det}N_{\a\b}}
\,.\label{H14b}\ee

Consider now the N=4 heterotic string in D=4.
The appropriate matrix N is
\be
N=S_{1}L+iS_{2}M^{-1}\;\;\;,\;\;\;S=S_1+iS_2
\,.\label{H15}\ee
Performing an overall duality as in (\ref{H9}) we obtain
\be
N'=-N^{-1}=-{S_{1}\over |S|^2}L+i{S_{2}\over |S|^2}M=-{S_{1}\over
|S|^2}L+i{S_{2}\over |S|^2}LM^{-1}L
\,.\label{H16}\ee
Thus, we observe that apart from an $S\to -1/S$ transformation on the
S field it also affects an O(6,22,$\Z)$ transformation by the matrix
$L$
which interchanges windings and momenta of the 6-torus.

The duality transformation which acts only on S is given by
$A=D=0$, $-B=C=L$ under which
\be
N'=-LN^{-1}L=-{S_{1}\over |S|^2}L+i{S_{2}\over |S|^2}M^{-1}
\,.\label{H17}\ee
The full SL(2,$\Z$) group acting on $S$ is generated by
\be
\left(\matrix{A&B\cr
C&D\cr}\right)=\left(\matrix{a\;{\bf 1}_{28}&b\; L\cr
c\;L&d\;{\bf 1}_{28}\cr}\right)\;\;\;,\;\;\;ad-bc=1
\,.\label{H18}\ee

Finally the duality transformation which acts as an O(6,22,$\Z$)
transformation
is given by $A=\Omega$, $D^{-1}=\Omega^{t}$, $B=C=0$.

\end{document}